\documentclass[12pt]{article}

.5 scaled \magstep4
\font\medio=cmr9.5 scaled \magstep2
\outer\def\beginsection#1\par{\medbreak\bigskip
      \message{#1}\leftline{\bf#1}\nobreak\medskip
\vskip-\parskip
      \noindent}
\usepackage{graphicx} 
\begin{document}
\bibliographystyle {unsrt}

\titlepage

\begin{flushright}
CERN-PH-TH/2011-176
\end{flushright}

\vspace{10mm}
\begin{center}
{\Large Magnetized $\Lambda$CDM inhomogeneities and the cosmic dark ages}\\ 
\vspace{1.5cm}
 Massimo Giovannini\footnote{Electronic address: massimo.giovannini@cern.ch}\\
\vspace{1.5cm}
{{\sl Department of Physics, 
Theory Division, CERN, 1211 Geneva 23, Switzerland }}\\
\vspace{0.5cm}
{{\sl INFN, Section of Milan-Bicocca, 20126 Milan, Italy}}\\
\vspace{0.5cm}
\end{center}

\vskip 0.5cm
\centerline{\medio  Abstract}
Exact solutions of the perturbations equations in the magnetized $\Lambda$CDM scenario 
are presented. They apply during the dark ages and, more specifically, after the baryons are freed from the drag of the photons.  The magnetized growth rate of matter perturbations is compared with the growth index obtained in the concordance paradigm and under the assumption that dark energy does not cluster for a redshift window ranging from the epoch of reionization to the stage of dark-energy dominance.  The constraints derived from this analysis 
of are shown to be qualitatively complementary and quantitatively competitive with the bounds stemming from the analysis of the distortion patterns induced by the magnetized adiabatic mode on the temperature and polarization anisotropies of the Cosmic Microwave Background.
\noindent

\vspace{5mm}

\vfill
\newpage
\renewcommand{\theequation}{1.\arabic{equation}}
\setcounter{equation}{0}
\section{Introduction} 
\label{sec1}
After the baryons are freed from the Compton drag of the photons\footnote{In what follows the values of the redshifts 
$z$ will be estimated in terms of the standard $\Lambda$CDM paradigm analyzed in the light 
of the WMAP 7yr data alone \cite{wmap7a,wmap7b,wmap7c}.} (i.e. $z_{\mathrm{drag}}  = 1020.3 \pm 1.4$) 
 large-scale magnetic fields contribute to the evolution of the inhomogeneities in the spatial curvature and in the matter distribution even if observational constraints on their effects are 
 lacking.  After the drag epoch (i.e. $z < z_{\mathrm{drag}}$) and 
before the Universe is reionized (i.e. $z \geq z_{\mathrm{re}}= 10.5 \pm 1.2$) the Universe 
enters a very interesting epoch (often dubbed as dark age) when the first gravitationally bound system are formed. 
The purpose of the present paper is to explore the evolution of large-scale magnetic fields during the 
dark ages in the minimal framework of the magnetized $\Lambda$CDM scenario  (m$\Lambda$CDM in what follows)
and in the simplest magnetohydrodynamical description which neglects the propagation of all the electromagnetic excitations of the plasma.

Probably the best evidence of reionization comes 
from the analysis of the large-scale temperature and polarization fluctuations in the Cosmic Microwave background (CMB).
Roughly speaking WMAP finds that about $8$\% of the CMB photons were scattered by free electrons in the intergalactic medium, but only $4$ \% could have been scattered  by the intergalactic medium for redshifts $z< 6$. This means that the Universe must have been already reionized for $z \geq {\mathcal O}(10)$. The direct estimates 
of $z_{\mathrm{re}}$ in terms of he WMAP 7yr data are consistent with  the 
Sloan Digital Sky Survey (SDSS) \cite{sdss1,sdss2} reporting the observations 
of quasars whose absorption spectra show, for $z> 6$, a substantial increase in the 
fraction of ionized hydrogen. In \cite{mg1,mg2} it has been argued, among other things, that the thickness of the last scattering surface and the optical depth at reionization are only mildly sensitive to the presence 
of an ambient magnetic field; the effect of reionization has then been parametrized, as in the 
conventional situation, with a second  Gaussian peak of the visibility function centered around the reionization 
time. Conversely the inhomogeneities of the geometry and of the sources are affected by the presence of the 
ambient magnetic field whose effects on the baryons and on the geometry can never be completely switched off.

Prior to photon decoupling the electrons, ions and photons are tightly coupled together; semi-analytic and numerical treatments of this problem are available in the presence of large-scale magnetic fields \cite{mg1,mg2}. Useful applications 
of the analytic treatments contemplate, for instance, explicit formulas for the distortions and for the shifts in the positions 
of the acoustic peaks as well as of the polarization peaks. For $z < z_{\mathrm{drag}}$ the evolution of the inhomonogeneities in the m$\Lambda$CDM scenario simplifies in many respects and the evolution
of the large-scale inhomogeneities can be solved exactly. The solutions for the evolution
 of the metric inhomogeneities in the m$\Lambda$CDM paradigm can also be extended to 
 complementary cases provided the dark energy does not cluster as it can be conveniently assumed for different applications. In the literature there exist some exact solutions 
of the evolution equations of metric perturbations in pressureless media like the ones of Hwang \cite{hwang1,hwang2} or the ones discussed in Refs. \cite{et1,et2,et3}. 
In the present paper after deriving the evolution equations of metric inhomogeneities for $z< z_{\mathrm{drag}}$, the exact solutions of the perturbation equations will be reported in the presence of a fully inhomogeneous magnetic field in the magnetohydrodynamical approximation stipulating, as customarily done, 
that the magnetic field, the Ohmic current and the electric field are all solenoidal (comoving) vectors. 

It is relevant to mention that various projects or radio arrays aim at a direct scrutiny of the dark ages like LOFAR (low-frequency array) \cite{lofar}, MWA (Murchison Wide-Field Array) \cite{mwa}, 
 the enhanced VLA (Very Large Array) \cite{VLA} and the SKA (square kilometer array) \cite{SKA}. One of the common strategies of these programs is the direct observational 
tomography of the (redshifted) $21$ cm emission. The logic is to construct radio arrays capable of mapping the 
three-dimensional distribution of primeval hydrogen. Such a spatial distribution with its structures, its clumpiness, possibly its polarization bears the mark of the evolution through the dark ages and, most notably, of the first galaxies and stars producing ultraviolet emission and hence holes in the distribution of the protohydrogen. The role 
of cosmic magnetism in this epoch should be more carefully explored but the growth of matter inhomogeneities during the dark ages is certainly affected by the presence of large-scale magnetic fields, as general physical considerations suggest and as the solutions presented hereunder confirm.
In section \ref{sec2} we are going  to present the evolution of the inhomogeneities in the 
m$\Lambda$CDM model and for $z< z_{\mathrm{drag}}$. In section \ref{sec3} the exact solutions of the 
system will be presented. In section \ref{sec4}  the exact solutions are applied to the explicit 
calculation of the magnetized growth rate which is then compared to the conventional result 
of the concordance scenario with the purpose of deriving en explicit bound valid over the 
inhomogeneity scales relevant to structure formation.

\renewcommand{\theequation}{2.\arabic{equation}}
\setcounter{equation}{0}
\section{Evolution equations in real space} 
\label{sec2}
The drag redshift corresponds, approximately, to the moment when the baryons are freed from the Compton 
drag of the photons and it can be expressed in terms of the present critical fraction of matter and of baryons \cite{drag1}:
\begin{eqnarray}
z_{\mathrm{drag}} &=& \frac{1291 \, \omega_{\mathrm{M}0}^{0.251}}{1 + 0.659\, 
\omega_{\mathrm{M}0}^{0.828}}[ 1 + q_{1} \omega_{\mathrm{b}0}^{q2}],
\label{TP8}\\
q_{1} &=& 0.313 \, \omega_{\mathrm{M}0}^{-0.419}[ 1 + 0.607 \, \omega_{\mathrm{M}0}^{0.674} ],
\qquad q_{2} = 0.238 \, \omega_{\mathrm{M}0}^{0.223},
\label{TP9}
\end{eqnarray}
where $\omega_{\mathrm{M}0} = h_{0}^2 \Omega_{\mathrm{M}0}$ and, more generally, 
$\omega_{x\, 0} = h_{0}^2 \Omega_{x0}$ for a generic species $x$. 
For the present purposes, it is practical to decompose the synchronous gauge fluctuations directly in real 
space as 
\begin{equation}
\nabla^2 \delta_{\mathrm{s}} g_{ij}(\vec{x}, \tau) = a^2(\tau) \biggl\{ \partial_{i} \partial_{j} h(\vec{x},\tau) + 6 \biggl[ \partial_{i} \partial_{j} \xi(\vec{x},\tau) - \frac{1}{3} \delta_{i j}  \nabla^2\xi(\vec{x},\tau)\biggr]\biggr\},
\label{SY0}
\end{equation}
where we have assumed a conformally flat background geometry characterized by the scale factor 
$a(\tau)$ as it happens in the case of the $\Lambda$CDM paradigm.  
The density contrasts of dark matter, baryons and dark energy will be denoted, respectively, by 
$\delta_{\mathrm{c}}$, $\delta_{\mathrm{b}}$ and $\delta_{\mathrm{de}}$. 
The Hamiltonian constraint stemming from the $(00)$ component of the perturbed Einstein equations in the gauge 
(\ref{SY0}) is\footnote{The standard notations for the Hubble rates will be adopted, i.e. $H = a {\mathcal H}$ 
with $a {\mathcal H} = \partial_{\tau} a$.}
\begin{equation}
2 \Xi + {\mathcal H} \partial_{\tau} h + 3 {\mathcal H}^2 \biggl\{
\Omega_{\mathrm{M}} \biggl[
\biggl(\frac{\omega_{\mathrm{b}0}}{\omega_{\mathrm{M}0}}\biggr) \delta_{\mathrm{b}} +
\biggl(\frac{\omega_{\mathrm{c}0}}{\omega_{\mathrm{M}0}}\biggr) \delta_{\mathrm{c}} \biggr]
+  \Omega_{\mathrm{de}} \delta_{\mathrm{de}}\biggr\}
+ \ell_{\mathrm{P}}^2 a^2 [\delta_{\mathrm{s}} \rho_{\mathrm{B}} + \delta_{\mathrm{s}} \rho_{\mathrm{E}}] =0,
\label{SY1}
\end{equation}
where $\Xi(\vec{x},\tau) = \nabla^2 \xi(\vec{x},\tau)$, $\ell_{\mathrm{P}} = \sqrt{8 \pi G}$ and 
\begin{eqnarray}
&& \delta_{\mathrm{s}} \rho_{\mathrm{B}} = \frac{B^2(\vec{x},\tau) }{8 \pi a^4}, \qquad 
\delta_{\mathrm{s}} p_{\mathrm{B}} =\frac{\delta_{\mathrm{s}} \rho_{\mathrm{B}}}{3}, \qquad 
F(\vec{x},\tau) = \frac{\vec{\nabla}\cdot(\vec{E} \times \vec{B})}{4 \pi a^4},
\nonumber\\
&&  \delta_{\mathrm{s}} \rho_{\mathrm{E}} = \frac{E^2(\vec{x},\tau) }{8 \pi a^4}, \qquad 
\delta_{\mathrm{s}} p_{\mathrm{E}} =\frac{\delta_{\mathrm{s}} \rho_{\mathrm{E}}}{3};
\label{SY1a}
\end{eqnarray}
$\vec{E}$ and $\vec{B}$ denote, respectively, the comoving electric and magnetic fields.
The momentum constraint is instead given by 
\begin{equation}
 \partial_{\tau} \Xi = \frac{3}{2} {\mathcal H}^2 \biggl\{ (1 + w_{\mathrm{de}})  \Omega_{\mathrm{de}} \theta_{\mathrm{de}}+ \Omega_{\mathrm{M}} \biggl[ \biggl(\frac{\omega_{\mathrm{b}0}}{\omega_{\mathrm{M}0}}\biggr) \theta_{\mathrm{b}} + \biggl(\frac{\omega_{\mathrm{c}0}}{\omega_{\mathrm{M}0}}\biggr) \theta_{\mathrm{c}}\biggr] \biggr\}
+ \frac{\ell_{\mathrm{P}}^2}{2} a^2 F(\vec{x},\tau).
\label{SY2}
\end{equation}
In the magnetohydrodynamical approximation adopted in the minimal m$\Lambda$CDM model the displacement current is neglected since the frequencies involved by the present considerations are very small in comparison with the plasma 
frequency and the typical length-scales much larger than the Debye length. In the latter approximation not only 
the comoving magnetic field, the comoving electric field and the comoving Ohmic current are 
all solenoidal, i.e.
\begin{equation}
\vec{\nabla} \cdot \vec{B} =0, \qquad \vec{\nabla} \cdot \vec{E}=0, \qquad 
\vec{\nabla}\cdot \vec{J} \equiv \frac{1}{4 \pi} \vec{\nabla} \cdot(\vec{\nabla}\times \vec{B}) =0.
\label{SY6}
\end{equation}
The solenoidal nature of the Ohmic current is consistent with the absence of the displacement current, 
as shown in the last relation of Eq. (\ref{SY6}) where $\vec{J}$ has been expressed, using the Maxwell equations for the 
comoving fields, as $\vec{\nabla}\times \vec{B}/(4\pi)$.  When the ionization fraction $x_{\mathrm{e}}$ drops almost suddenly from $1$ to about
$10^{-5} \omega_{\mathrm{M}0}/\omega_{\mathrm{b}0}$ 
the concentration of the free 
charge carriers diminishes but the drop of the free charge carriers is faster than the drop of the temperature so that 
the plasma parameter decreases 
\begin{equation}
g_{\mathrm{plasma}} = \frac{1}{V_{\mathrm{D}} n_{0} x_{\mathrm{e}}} = 24 e^{3}  \sqrt{\frac{\zeta(3)}{\pi}} 
\sqrt{x_{\mathrm{e}} \eta_{\mathrm{b}0}}
= 2.3\times 10^{-7} \sqrt{x_{\mathrm{e}}} \biggl(\frac{h_{0}^2\Omega_{\mathrm{b0}}}{0.02258}\biggr)^{1/2},
\label{con4}
\end{equation}
where $V_{\mathrm{D}} = 4 \pi \lambda_{\mathrm{D}}^3/3$ is the volume of the Debye sphere, 
$\lambda_{\mathrm{D}}$ is the Debye length, $\eta_{\mathrm{b}0}$ is the ratio between the baryon and the photon concentrations and $\zeta(3)=1.202$. The smallness 
of $g_{\mathrm{plasma}}$ guarantees the validity of the plasma approximation; moreover, since the 
masses of the charge carriers are much larger than the kinetic temperature of the corresponding species, 
 the conductivity $\sigma$ is proportional to the inverse of $g_{\mathrm{plasma}}$. This means that 
 the electric fields will be suppressed as inverse powers of the conductivity or as positive powers 
 of the plasma parameter. Thus the electric fields can be dropped from the dynamical equations for 
$h$ and $\Xi$:
\begin{eqnarray}
&& \partial_{\tau}^2 h + 2 {\mathcal H} \partial_{\tau} h + 2 \Xi =  9 {\mathcal H}^2 (1 - \Omega_{\mathrm{M}}) 
w_{\mathrm{de}}\delta_{\mathrm{de}}  +  3\ell_{\mathrm{P}}^2 a^2 \,  \delta_{\mathrm{s}} p_{\mathrm{B}},
\label{SY4}\\
&& \partial_{\tau}^2 \Xi + 2 {\mathcal H} \partial_{\tau}\Xi = \frac{\ell_{\mathrm{P}}^2}{2 a^2} 
\vec{\nabla}\cdot(\vec{J} \times\vec{B}),
\label{SY5}
\end{eqnarray}
where $w_{\mathrm{de}}$ denotes the barotropic index for dark energy. The evolution equation for the baryons can be written as 
\begin{equation}
\partial_{\tau} \theta_{\mathrm{b}} + {\mathcal H} \theta_{\mathrm{b}} = \frac{\vec{\nabla}\cdot (\vec{J} \times \vec{B})}{a^4 \rho_{\mathrm{b}}}, \qquad 
\partial_{\tau} \delta_{\mathrm{b}} = \frac{1}{2}\partial_{\tau}h - \theta_{\mathrm{b}},
\label{B1}
\end{equation}
where $\rho_{\mathrm{b}}$ is the baryonic mass density.  The corresponding equations for the dark-matter species are:
\begin{equation}
\partial_{\tau}\theta_{\mathrm{c}} + {\mathcal H} \theta_{\mathrm{c}} =0, \qquad 
\partial_{\tau}\delta_{\mathrm{c}}= \frac{1}{2} \partial_{\tau} h- \theta_{\mathrm{c}}.
\label{B2}
\end{equation}
The evolution equations for the dark energy are instead given by:
\begin{eqnarray}
&& \partial_{\tau} \theta_{\mathrm{de}} + {\mathcal H} (1 - 3 c_{\mathrm{de}}^2) \theta_{\mathrm{de}} + 
\frac{c_{\mathrm{de}}^2}{w_{\mathrm{de}} +1} \Delta_{\mathrm{de}} =0,
\label{B3}\\
&& \partial_{\tau} \Delta_{\mathrm{de}} + 3 {\mathcal H}(c_{\mathrm{de}}^2 - w_{\mathrm{de}}) \Delta_{\mathrm{de}} = (w_{\mathrm{de}} +1)\biggl[ \Theta_{\mathrm{de}} + 9 {\mathcal H}^2 (c_{\mathrm{de}}^2 - w_{\mathrm{de}}) \theta_{\mathrm{de}} - \frac{\nabla^2 \partial_{\tau} h}{2}\biggr],
\label{B4}
\end{eqnarray}
where $\Delta_{\mathrm{de}} = \nabla^2 \delta_{\mathrm{de}}$ and $\Theta_{\mathrm{de}} = 
\nabla^2 \theta_{\mathrm{de}}$; $c_{\mathrm{de}}$ denotes the sound speed of dark energy which is assigned 
independently of the barotropic index.

\renewcommand{\theequation}{3.\arabic{equation}}
\setcounter{equation}{0}
\section{Analytic solutions of system} 
\label{sec3}
It is practical to start the integration of the system from the evolution equation of the baryon 
velocity reported in Eq. (\ref{B1}). By using, as integration variable, $\alpha= a/a_{\mathrm{de}}$, the result of the integration o Eq. (\ref{B1}) can be expressed in terms of conventional hypergeometric functions \cite{abr,tric}:
\begin{eqnarray}
\theta_{\mathrm{b}}(\vec{x},\alpha) &=& \theta_{\mathrm{b}}(\vec{x},\alpha_{*})\biggl(\frac{\alpha_{*}}{\alpha}\biggr) 
+ {\mathcal M}_{\theta}(w_{\mathrm{de}}, z_{\mathrm{de}}, \Omega_{\mathrm{M}0}) \,{\mathcal S}_{\mathrm{B}}(\vec{x}, \alpha) \, {\mathcal G}_{\theta}(\alpha,w_{\mathrm{de}}),
\nonumber\\
 {\mathcal M}_{\theta}(w_{\mathrm{de}}, z_{\mathrm{de}}, \Omega_{\mathrm{M}0})&=& \frac{2 ( z_{\mathrm{de}} +1)^{7/2}}{(3 w_{\mathrm{de}} +1 ) \, H_{0} \, 
 \sqrt{\Omega_{\mathrm{M}0}}}, \qquad {\mathcal S}_{\mathrm{B}}(\vec{x}, \alpha) = 
 \frac{\vec{\nabla} \cdot(\vec{J} \times \vec{B})}{\rho_{\mathrm{b}}(\alpha)},
\nonumber\\
{\mathcal G}_{\theta}(\alpha,w_{\mathrm{de}}) &=& \alpha^{( 3 w_{\mathrm{de}} - 7)/2}\,  
F\biggl[\frac{1}{2} + \frac{1}{6 w_{\mathrm{de}}},\, \frac{1}{2};\, \frac{3}{2} + \frac{1}{6 w_{\mathrm{de}}}; - \alpha^{3 w_{\mathrm{de}}}\biggr],
\label{SOL1}
\end{eqnarray}
where $\alpha_{*}$ and $\theta_{\mathrm{b}}(\vec{x},\alpha_{*})$ denote, respectively,  the initial integration time and the 
corresponding (inhomogeneous) value of the velocity field at $\alpha_{*}$; 
$F[a,\, b;\, c;\, z] = \,\,_{2}F_{1}[a,\, b;\, c;\, z]$ denotes the hypergeometric function \cite{abr,tric}. 
The scale factor itself (appearing, for instance, 
in Eq. (\ref{SY0}), is normalized to $1$ at the 
present epoch (i.e. $a_{0} =1$) so that the following relations hold between $a_{\mathrm{de}}$ 
and the corresponding critical fractions of matter and dark energy:
\begin{equation}
a_{\mathrm{de}} = \frac{1}{z_{\mathrm{de}} +1} = \biggl(\frac{\Omega_{\mathrm{de}0}}{\Omega_{\mathrm{M}0}}\biggr)^{\frac{1}{3 w_{\mathrm{de}}}},\qquad \Omega_{\mathrm{de}0} = 1 - \Omega_{\mathrm{M}0},\qquad \Omega_{\mathrm{M}0} = \Omega_{\mathrm{c}0} + \Omega_{\mathrm{b}0}.
\label{SOL1A}
\end{equation}
It is well known \cite{PV1} that, in the synchronous gauge, the coordinate 
 system is only partially fixed. The remaining gauge freedom must be removed from the initial conditions 
 to avoid the potentially dangerous presence of spurious (i.e. gauge) 
 modes. This problem is handled by removing the 
 remaining gauge freedom or by expressing the results in terms of appropriate gauge-invariant 
 variables \cite{cosmics2,bardeen} (see also \cite{weinberg}).
It is customary to select the dark matter rest frame and choose $\theta_{\mathrm{c}}(\vec{x}, \alpha) =0$ to completely 
fix the coordinate system. In the latter case Eq. (\ref{SY2}) reads:
\begin{equation} 
\frac{\partial \Xi}{\partial\alpha} = \frac{3 {\mathcal H}}{2\, \alpha} \biggl[\biggl(\frac{\omega_{\mathrm{b}0}}{\omega_{\mathrm{M}0}}\biggr)\Omega_{\mathrm{M}}\, \theta_{\mathrm{b}} + (1 + w_{\mathrm{de}}) \Omega_{\mathrm{de}} \theta_{\mathrm{de}} \biggr],
\label{SOL2}
\end{equation}
where $\Omega_{\mathrm{de}} = 1 - \Omega_{\mathrm{M}}$ and the contribution of the radiation has 
been neglected since the drag redshift occurs when the background is already dominated by matter. 
If the dark energy is incompressible, its fluctuations can be neglected (i.e. $\theta_{\mathrm{de}} =0$ and $\delta_{\mathrm{de}} =0$) but the contribution on the background obviously persists since 
\begin{equation}
{\mathcal H}(\alpha) = H_{0}\,\, \frac{\sqrt{\Omega_{\mathrm{M}0} }}{\sqrt{\alpha} \, \sqrt{\Omega_{\mathrm{M}}(\alpha)}} \sqrt{z_{\mathrm{de}}+1} , \qquad \Omega_{\mathrm{M}}(\alpha) = \frac{\alpha^{3 w_{\mathrm{de}}}}{\alpha^{3 w_{\mathrm{de}}} + 1}.
\label{SOL3}
\end{equation}
Note that the relations of Eq. (\ref{SOL3}) can be used for a swift derivation of Eq. (\ref{SOL1}).  Dark energy is 
incompressible in the context of the m$\Lambda$CDM scenario (where 
$w_{\mathrm{de}} \to -1$ and, according to Eq. (\ref{B4}), $\delta_{\mathrm{de}} =0$ exactly) but also in all the situations where $w_{\mathrm{de}}$ can be considered close to $-1$ (where, approximately, $\delta_{\mathrm{de}} \simeq 0$). If the dark energy is compressible (i.e. $\theta_{\mathrm{de}} \neq 0$)  then $c_{\mathrm{de}}^2$ must be different from $w_{\mathrm{de}}$ since $w_{\mathrm{de}} \leq -1/3$. The bounds on $c_{\mathrm{de}}^2$ are currently rather loose and we shall assume, as customarily done \cite{cs1}, that $0\leq c_{\mathrm{de}}^2 \leq 1$ in the context of the so-called $w$CDM 
scenario\footnote{By analyzing different data sets in the light of the $w$CDM scenario,  the error bars on $w_{\mathrm{de}}$ either increase or they are restricted to an interval pinning down the $\Lambda$CDM value $w_{\mathrm{de}} = -1$. Values $w_{\mathrm{de}} < -1$ will 
be excluded since, in these cases, the background may evolve towards a singularity in the future.
This is only a practical choice since, in principle the present discussion can also be extended to the 
situation $w_{\mathrm{de}} < -1$}. Since the  barotropic index $w_{\mathrm{de}}$ and the sound speed $c_{\mathrm{de}}$ are assigned indipendently the total pressure fluctuation inherits a non-adiabatic contribution which is proportional 
to $(w_{\mathrm{de}} - c_{\mathrm{de}}^2)$ \cite{cs2} and this is the rationale of the appearance of such a term 
in Eq. (\ref{B4}). For sake of simplicity we shall choose $c_{\mathrm{de}}=0$ and, in this case, Eq. (\ref{SOL2}) 
can be directly integrated and the result is: 
\begin{eqnarray}
\Xi(\vec{x},\alpha) &=& \Xi(\vec{x},\alpha_{*}) + {\mathcal N}_{\xi}(\vec{x}, z_{\mathrm{de}}, w_{\mathrm{de}}) {\mathcal F}_{1}(\alpha, w_{\mathrm{de}})
\nonumber\\
&+& {\mathcal Q}_{\xi}(z_{\mathrm{de}}, w_{\mathrm{de}}) {\mathcal S}_{\mathrm{B}}(\vec{x},\alpha) {\mathcal F}_{2}(\alpha, w_{\mathrm{de}}) - {\mathcal P}_{\xi}(\vec{x}, z_{\mathrm{de}}, w_{\mathrm{de}}) {\mathcal F}_{3}(\alpha, w_{\mathrm{de}}),
\label{SOL4}
\end{eqnarray}
where ${\mathcal N}_{\xi}(\vec{x}, z_{\mathrm{de}}, w_{\mathrm{de}}) $, ${\mathcal Q}_{\xi}(z_{\mathrm{de}}, w_{\mathrm{de}})$ and ${\mathcal P}_{\xi}(\vec{x}, z_{\mathrm{de}}, w_{\mathrm{de}})$ 
are defined as:
\begin{eqnarray}
{\mathcal N}_{\xi}(\vec{x}, z_{\mathrm{de}}, w_{\mathrm{de}})  &=& \frac{H_{0} \, \sqrt{\Omega_{\mathrm{M}0}}}{(w_{\mathrm{de}} -1)} 
\sqrt{z_{\mathrm{de}}+1} \biggl(\frac{\omega_{\mathrm{b}0}}{\omega_{\mathrm{M}0}}\biggr) 
\alpha_{*} \theta_{\mathrm{b}}(\vec{x}, \alpha_{*}),
\nonumber\\
 {\mathcal Q}_{\xi}(z_{\mathrm{de}}, w_{\mathrm{de}}) &=& \frac{3}{9 w_{\mathrm{de}}^2 -1} 
\biggl(\frac{\omega_{\mathrm{b}0}}{\omega_{\mathrm{M}0}}\biggr)  ( z_{\mathrm{de}}+1)^4
\nonumber\\
{\mathcal P}_{\xi}(\vec{x}, z_{\mathrm{de}}, w_{\mathrm{de}}) &=& H_{0}\,\sqrt{\Omega_{\mathrm{M}0}} \, 
\alpha_{*} \, \theta_{\mathrm{de}}(\vec{x},\alpha_{*})\, \sqrt{z_{\mathrm{de}} +1}.
\label{SOL4a}
\end{eqnarray}
The functions ${\mathcal F}_{i}(\alpha, w_{\mathrm{de}})$ (with $i= 1,\,... 3$) are either conventional or generalized hypergeometric functions \cite{abr,tric}: 
\begin{eqnarray}
{\mathcal F}_{1}(\alpha, w_{\mathrm{de}}) &=& \alpha^{3(w_{\mathrm{de}} -1)/2} 
F\biggl[ \frac{1}{2} - \frac{1}{2 w_{\mathrm{de}}}, \frac{1}{2}; \frac{3}{2} - \frac{1}{2 w_{\mathrm{de}}}; 
- \alpha^{3 w_{\mathrm{de}}}\biggr],
\nonumber\\
{\mathcal F}_{2}(\alpha, w_{\mathrm{de}}) &=& \alpha^{(3 w_{\mathrm{de}}-4)}\,\,_{3} F_{2}\biggl[a_{1},\, a_{2},\, a_{3}\,; b_{1},\, b_{2}; - \alpha^{ 3 w_{\mathrm{de}}}\biggr],
\nonumber\\
{\mathcal F}_{3}(\alpha, w_{\mathrm{de}}) &=& \alpha^{- 3( 1 + w_{\mathrm{de}})/2} F\biggl[ \frac{1}{2},\, - \frac{1}{2} - \frac{1}{2 w_{\mathrm{de}}}; \frac{1}{2} - \frac{1}{2 w_{\mathrm{de}}}; - \alpha^{ 3 w_{\mathrm{de}}}\biggr],
\label{SOL4b}
\end{eqnarray}
where  $_{p}F_{q}[a_{1},\,...,\,a_{p}; b_{1},\,...,\, b_{q}; z]$ denotes the generalized hypergeometric function \cite{abr,tric} with: 
\begin{eqnarray}
a_{1}&=&1,\qquad  a_{2} = 1 - \frac{1}{3 w_{\mathrm{de}}},\qquad a_{3} = 1 + \frac{1}{6 w_{\mathrm{de}}},
\nonumber\\
 b_{1} &=& 2 - \frac{1}{3 w_{\mathrm{de}}}, \qquad 
b_{2} = \frac{3}{2} + \frac{1}{6 w_{\mathrm{de}}}.
\label{SOL4c}
\end{eqnarray}
Concerning the results of Eqs. (\ref{SOL4}), (\ref{SOL4a}) and (\ref{SOL4b}) few technical comments are on order.
The integrals required for the actual solution can be performed by making extensive use of the 
transformation formulas of the hypergeometric functions \cite{abr}. A particularly useful 
formula is the one stipulating that 
\begin{equation}
F[a,\, b,\,; c;\, - \beta^{3 w_{\mathrm{de}}}] = \biggl(1 + \beta^{3 w_{\mathrm{de}}}\biggr)^{c - a - b} F[c-a,\, c- b,\,; c;\, - \beta^{3 w_{\mathrm{de}}}].
\label{TR1}
\end{equation}
The repeated use of Eq. (\ref{TR1}) simplifies various expressions which can be reduced, in some cases, to the following 
indefinite integral
\begin{eqnarray}
&& \int \, \beta^{m} F\biggl[ 1,\, 1+ n;\, \frac{3}{2} + n;\, - \beta^{3 s}\biggr] \, \, d\beta = \frac{\beta^{m + 1}}{m + 1} \,\, _{3}F_{2}[a_{1},\,a_{2}; b_{1},\,b_{2},\, b_{3}; - \beta^{ 3 s}],
\nonumber\\
&& a_{1} = 1,\qquad a_{2} = 1 + n,\qquad a_{3}= \frac{m + 1}{ 3 s},\qquad b_{1} = \frac{3}{2} + n,\qquad b_{2} = 1 + \frac{m+1}{3 s},
\nonumber
\end{eqnarray}
which is a consequence of the formula giving the nth derivative of a generalized hypergeometric function \cite{nist}.
The solution of Eqs. (\ref{SOL4})  and (\ref{SOL4a})--(\ref{SOL4c})
solves also Eq. (\ref{SY5}) whose explicit form in terms of the notation used in this section is:
\begin{equation}
\frac{\partial^2 \Xi}{\partial\alpha^2} + \frac{1}{\alpha} \biggl[ \frac{5}{2} - \frac{3}{2} w_{\mathrm{de}} ( 1 - \Omega_{\mathrm{M}}) \biggr] \frac{\partial \Xi}{\partial \alpha}= \frac{\ell_{\mathrm{P}}^2\, (z_{\mathrm{de}} +1)^2}{2 \alpha^4 {\mathcal H}^2}  \vec{\nabla}\cdot (\vec{J} \times\vec{B}).
\label{SOL6}
\end{equation}
By knowing explicitly the solution in terms of $\Xi(\vec{x}, \alpha)$ it is immediate to compute 
the curvature perturbations on comoving orthogonal hypersurfaces whose explicit expression 
in terms of $\xi$ can be written as:
\begin{equation}
{\mathcal R}(\vec{x}, \tau) = \xi(\vec{x},\tau) + \frac{{\mathcal H} \partial_{\tau} \xi(\vec{x},\tau)}{{\mathcal H}^2 - \partial_{\tau} {\mathcal H}}, 
\label{SOL7}
\end{equation}
which also implies that 
\begin{equation}
\nabla^2 {\mathcal R} = \Xi + \frac{2 \, \alpha}{3 [ 1 + w_{\mathrm{de}} ( 1 - \Omega_{\mathrm{M}}(\alpha))]}  
\frac{\partial \Xi}{\partial\alpha}.
\label{SOL8}
\end{equation}
Since the curvature perturbations are invariant for infinitesimal gauge transformations 
they will have the same value in any gauge. 
In the CDM rest frame,  adopted throughout to get rid of spurious gauge modes, $\delta_{\mathrm{c}}(\vec{x},\alpha)$ can be immediately computed
\begin{equation}
\delta_{\mathrm{c}}(\vec{x},\alpha) = \delta_{\mathrm{c}}(\vec{x},\alpha_{*}) + \frac{h(\vec{x},\alpha)}{2}.
\label{SOLDC}
\end{equation}
Similarly the solution for $\delta_{\mathrm{b}}(\vec{x},\alpha)$ turns out to be 
\begin{eqnarray}
\delta_{\mathrm{b}}(\vec{x},\alpha) &=& \delta_{\mathrm{b}}(\vec{x},\alpha_{*}) + \frac{h(\vec{x},\alpha)}{2} 
\nonumber\\
&-& \frac{2}{(3 w_{\mathrm{de}} -1)} \frac{ \alpha_{*} \theta_{\mathrm{b}}(\vec{x},\alpha_{*})}{H_{0} \sqrt{\Omega_{\mathrm{M}0}} \sqrt{z_{\mathrm{de}} +1}} \alpha^{( 3 w_{\mathrm{de}} -1)/2} F\biggl[ \frac{1}{2},\, \frac{1}{2} - \frac{1}{6 w_{\mathrm{de}}};\, \frac{3}{2} -\frac{1}{6 w_{\mathrm{de}}}; - \alpha^{3 w_{\mathrm{de}}} \biggr]
\nonumber\\
&-& \frac{4}{3 w_{\mathrm{de}}} \frac{(z_{\mathrm{de}}+1)^3}{H_{0}^2 \Omega_{\mathrm{M}0}} \, \frac{{\mathcal S}_{\mathrm{B}}(\vec{x},\alpha)}{\alpha^3} \biggl\{ F\biggl[ - \frac{1}{2},\, \frac{1}{6 w_{\mathrm{de}}};\, \frac{1}{2} + \frac{1}{6 w_{\mathrm{de}}},\, - \alpha^{3 w_{\mathrm{de}}} \biggr] -1\biggr\}.
\label{SOLDB}
\end{eqnarray}
The evolution equations of the the total matter density contrast $\delta_{\mathrm{m}}(\vec{x},\alpha)$
\begin{equation}
\delta_{\mathrm{m}} = \frac{\omega_{\mathrm{b}0}}{\omega_{\mathrm{M}0}}\delta_{\mathrm{b}}(\vec{x},\alpha) + 
 \frac{\omega_{\mathrm{c}0}}{\omega_{\mathrm{M}0}}\delta_{\mathrm{c}}(\vec{x},\alpha),
\nonumber
\end{equation}
can be derived by combining the previous equations. Defining 
the differential operator
\begin{equation}
{\mathcal L}_{1}(\alpha, w_{\mathrm{de}}, \Omega_{\mathrm{M}}) = \frac{\partial^{2} }{\partial\alpha^2} + 
\frac{3}{2 \alpha} [ 1 - w_{\mathrm{de}} ( 1 - \Omega_{\mathrm{M}})] \frac{\partial }{\partial \alpha} - \frac{3 \, \Omega_{\mathrm{M}}}{2\, \alpha^2},
\label{diffop1}
\end{equation}
the evolution of $\delta_{\mathrm{m}}(\vec{x},\alpha)$ is given by:
\begin{eqnarray}
&& {\mathcal L}^{(1)}(\alpha,w_{\mathrm{de}}, \Omega_{\mathrm{M}})\delta_{\mathrm{m}}(\vec{x},\alpha) = {\mathcal T}^{(1)}_{\mathrm{B}}(\vec{x}, \alpha), 
\label{SOLM1}\\
&& {\mathcal T}^{(1)}_{\mathrm{B}}(\vec{x}, \alpha) = - 
\frac{\omega_{\mathrm{b}0}}{\omega_{\mathrm{M}0}} 
\frac{(z_{\mathrm{de}}+1)^4}{\alpha^6 \,\, {\mathcal H}^2} {\mathcal S}_{\mathrm{B}}(\vec{x},\alpha) + 
\frac{\ell_{\mathrm{P}}^2}{8 \pi} \frac{(z_{\mathrm{de}}+1)^2}{\alpha^4 {\mathcal H}^2} B^2.
\label{SOLM2}
\end{eqnarray}
Note that in Eqs. (\ref{SOLM1}) and (\ref{SOLM2}) the dark energy component has been assumed to be smooth and 
incompressible. As noted before this assumption holds exactly when $w_{\mathrm{de}}=-1$ (as in the m$\Lambda$CDM case) and approximately when $w_{\mathrm{de}} \simeq -1$. 
The solution of Eqs. (\ref{SOLM1})--(\ref{SOLM2}) can be written as 
\begin{equation}
\delta_{\mathrm{m}}(\vec{x},\alpha)= {\mathcal C}_{1}(\vec{x},\alpha_{*}) {\mathcal F}_{4}(\alpha, w_{\mathrm{de}}) + 
{\mathcal C}_{2}(\vec{x},\alpha_{*}) {\mathcal F}_{5}(\alpha, w_{\mathrm{de}}) 
+ \int_{\alpha_{*}}^{\alpha}  {\mathcal T}^{(1)}_{\mathrm{B}}(\vec{x}, \beta) G(\alpha,\,\beta,\, w_{\mathrm{de}}) \, d\beta, 
\label{SOLM3}
\end{equation}
where 
\begin{equation}
G^{(1)}(\alpha,\,\beta,\, w_{\mathrm{de}})  = \frac{{\mathcal F}_{4}(\beta,w_{\mathrm{de}}){\mathcal F}_{5}(\alpha, w_{\mathrm{de}}) - {\mathcal F}_{5}(\beta,w_{\mathrm{de}})
{\mathcal F}_{4}(\alpha, w_{\mathrm{de}})}{W(\beta, w_{\mathrm{de}})},
\label{SOLM3a}
\end{equation}
and $W(\beta,w_{\mathrm{de}})$ 
\begin{equation}
W(\beta,w_{\mathrm{de}}) = {\mathcal F}_{4}(\beta, w_{\mathrm{de}}) \partial_{\beta} {\mathcal F}_{5}(\beta, w_{\mathrm{de}}) -
{\mathcal F}_{5 }(\beta, w_{\mathrm{de}})\partial_{\beta}{\mathcal F}_{4}(\beta, w_{\mathrm{de}}),
\label{SOLM3b}
\end{equation}
denotes the  Wronskian of the two independent solutions of the homogeneous equation whose explicit expressions are
\begin{eqnarray}
{\mathcal F}_{4}(\alpha, w_{\mathrm{de}}) &=&
\alpha^{(3 w_{\mathrm{de}}-1)/2}  F\biggl[ \frac{1}{2} - \frac{1}{2 w_{\mathrm{de}}}, \frac{1}{2} + \frac{1}{3 w_{\mathrm{de}}} ;
\frac{3}{2} - \frac{1}{6 w_{\mathrm{de}}}; - \alpha^{3 w_{\mathrm{de}}}\biggr]
\nonumber\\
{\mathcal F}_{5}(\alpha, w_{\mathrm{de}})  &=& F\biggl[ - \frac{1}{3 w_{\mathrm{de}}}, \frac{1}{2 w_{\mathrm{de}}}; 
\frac{1}{2} + \frac{1}{6 w_{\mathrm{de}}}; - \alpha^{3 w_{\mathrm{de}}}\biggr].
\end{eqnarray}
Finally the solution for $h(\vec{x},\alpha)$ is given by 
\begin{eqnarray}
h(\vec{x},\alpha) &=& h(\vec{x}, \alpha_{*}) + 2 \delta_{\mathrm{m}}(\vec{x},\alpha) +
\nonumber\\
&+& {\mathcal N}_{h}(\vec{x},z_{\mathrm{de}},w_{\mathrm{de}}) {\mathcal F}_{5}(\alpha, w_{\mathrm{de}}) + {\mathcal Q}_{h}(z_{\mathrm{de}}, w_{\mathrm{de}})\, {\mathcal S}_{\mathrm{B}}(\vec{x},\alpha) \, {\mathcal F}_{6}(\alpha, w_{\mathrm{de}}),
\nonumber\\
{\mathcal N}_{h}(\vec{x},z_{\mathrm{de}},w_{\mathrm{de}}) &=& \frac{4\, \alpha_{*} \theta_{\mathrm{b}}(k,\alpha_{*})}{(3 w_{\mathrm{de}} -1) \, H_{0} \, \sqrt{\Omega_{\mathrm{M}0}} \, \sqrt{z_{\mathrm{de}} +1}} \biggl(\frac{\omega_{\mathrm{b}0}}{\omega_{\mathrm{M}0}}\biggr),
\nonumber\\
{\mathcal Q}_{h}(z_{\mathrm{de}}, w_{\mathrm{de}}) &=& \frac{4 ( z_{\mathrm{de}}+1)^{3}}{3 w_{\mathrm{de}}( 3 w_{\mathrm{de}} -1) H_{0}^2 \Omega_{\mathrm{M}0}} \biggl(\frac{\omega_{\mathrm{b}0}}{\omega_{\mathrm{M}0}}\biggr),
\nonumber\\
{\mathcal F}_{6}(\alpha, w_{\mathrm{de}}) &=& \alpha^{(3 w_{\mathrm{de}} -1)/2} \, F\biggl[\frac{1}{2},\,\frac{1}{2} - \frac{1}{6 w_{\mathrm{de}}},\, \frac{3}{2} - \frac{1}{6 w_{\mathrm{de}}}, - \alpha^{3 w_{\mathrm{de}}}\biggr],
\nonumber\\
{\mathcal F}_{7}(\alpha, w_{\mathrm{de}}) &=& \alpha^{3(w_{\mathrm{de}}-7)/2}\,\, F\biggl[1,\, 1 + \frac{1}{6 w_{\mathrm{de}}};\, \frac{3}{2} + \frac{1}{6 w_{\mathrm{de}}}; - \alpha^{ 3 w_{\mathrm{de}}}\biggr].
\end{eqnarray}
To normalize properly the solutions  
obtained for $\alpha > \alpha_{*}$ it is useful, even if not mandatory, to compute analytically the solutions 
for $\alpha < \alpha_{*}$ by assuming that the dark energy background is 
negligible. In this case the matter-radiation system can also be solved in the 
presence of large-scale magnetic fields and we shall be interested in 
computing the obtained solutions in the limit $ a_{*} \leq a \ll a_{\mathrm{eq}}$.
Defining, in analogy with Eq. (\ref{diffop1}), 
the differential operator ${\mathcal L}_{2}(\alpha)$
\begin{equation}
{\mathcal L}^{(2)}(\alpha) = \frac{\partial^2}{\partial \alpha^2} \, + 
\frac{3 \alpha + 2}{2 \alpha ( \alpha +1)} \frac{\partial}{\partial\alpha} - 
\frac{3}{2 \alpha (\alpha +1)}, \qquad \alpha= \frac{a}{a_{\mathrm{eq}}},
\label{SR1}
\end{equation}
the evolution of $\delta_{\mathrm{m}}(\vec{x},\alpha)$ in the range $ a_{*} \leq a \ll a_{\mathrm{eq}}$ can be written as:
\begin{equation}
{\mathcal L}^{(2)}(\alpha) \delta_{\mathrm{m}}(\vec{x},\alpha) = {\mathcal T}^{(2)}_{\mathrm{B}}(\vec{x},\alpha)
\label{SR2}
\end{equation}
where now the source term takes the form
\begin{equation}
{\mathcal T}^{(2)}_{\mathrm{B}}(\vec{x},\alpha) = 
\frac{1}{\alpha^2 (\alpha+1)}
\frac{\ell_{\mathrm{P}}^2 \, B^2\, (z_{\mathrm{eq}} + 1)^{3}}{8\pi H_{0}^2 \Omega_{\mathrm{M}0}} - \frac{1}{\alpha^4 (\alpha+1)}\biggl(\frac{\omega_{\mathrm{b}0}}{\omega_{\mathrm{M}0}}\biggr) \frac{(z_{\mathrm{eq}}+1)^3}{H_{0}^2 \Omega_{\mathrm{M}0}}\, {\mathcal S}_{\mathrm{B}}(\vec{x},\alpha).
\label{SR3}
\end{equation}
The solution of the previous equation reads
\begin{equation}
\delta_{\mathrm{m}}(\vec{x},\alpha) = {\mathcal D}_{1}(\vec{x}) 
{\mathcal F}_{8}(\alpha) + {\mathcal D}_{2}(\vec{x}) {\mathcal F}_{9}(\alpha)
+ \int_{\alpha_{\mathrm{i}}}^{\alpha}  {\mathcal T}^{(2)}_{\mathrm{B}}(\vec{x}, \beta) G^{(2)}(\alpha,\,\beta) \, d\beta,
\label{SR4}
\end{equation}
where $\alpha_{\mathrm{i}} = a_{\mathrm{i}}/a_{\mathrm{eq}}$ denotes the initial 
reference value of the scale factor while $G^{(2)}(\alpha,\,\beta,)$ and $W(\beta)$ are defined as
\begin{equation}
G^{(2)}(\alpha,\,\beta,)  = \frac{{\mathcal F}_{8}(\beta){\mathcal F}_{9}(\alpha) - {\mathcal F}_{9}(\beta)
{\mathcal F}_{8}(\alpha)}{W(\beta)},\qquad  W(\beta, w_{\mathrm{de}})=\frac{1}{\beta \sqrt{\beta+1}}.
\label{SR5}
\end{equation}
The two linearly independent solutions ${\mathcal F}_{8}(\alpha)$ and ${\mathcal F}_{9}(\alpha)$ are given, this time, by
the following expressions:
\begin{eqnarray}
&& {\mathcal F}_{8}(\alpha) = 1 + \frac{3}{2} \alpha, \qquad \alpha = \frac{a}{a_{\mathrm{eq}}},
\nonumber\\
&&{\mathcal F}_{9}(\alpha) = - \biggl(1 + \frac{3}{2} \alpha\biggr) \ln{\biggl[\frac{\sqrt{\alpha+1}+1}{\sqrt{\alpha + 1} -1}\biggr]} + 3 \sqrt{\alpha +1}.
\label{SR6}
\end{eqnarray}
Following the remarks already proposed around Eq. (\ref{SOL7}) it is 
useful to reiterate that the gauge dependent description pursued 
here can be complemented by appropriate gauge-invariant treatments. 
For instance  the variables 
\begin{equation}
\zeta_{\mathrm{c}} = \xi + \frac{\delta_{\mathrm{c}}}{3}, \qquad 
\zeta_{\mathrm{b}} = \xi + \frac{\delta_{\mathrm{b}}}{3},\qquad 
\zeta_{\mathrm{m}} = \xi + \frac{\delta_{\mathrm{m}}}{3},
\label{GI}
\end{equation}
are gauge-invariant 
and become, in the uniform curvature gauge \cite{hwang1,hwang2}, the density contrasts of the single species. In equally correct terms we could also argue 
that the $\zeta_{x}$ of the species $x$ measures the curvature perturbations 
on the hypersurfaces where the energy density of $x$ is uniform. From the perspective of the synchronous gauge the latter statements can be appreciated 
by noticing that 
$\zeta$ is proportional to $\xi$ which is related to ${\mathcal R}$ 
(see Eq. (\ref{SOL7})) corresponding, in turn, to the curvature perturbation in the comoving orthogonal gauge.

The results derived in the present section can be used for different purposes. In what follows the attention will be focused on the calculation of the growth factor of matter inhomogeneities and on its comparison with the results obtained in the more conventional situation where magnetic fields are assumed to be absent. 
\renewcommand{\theequation}{4.\arabic{equation}}
\setcounter{equation}{0}
\section{Applications and discussions} 
\label{sec4}
The integral appearing in Eq. (\ref{SR4}) can be evaluated explicitly and 
the obtained results expanded in the limit 
$\alpha_{\mathrm{i}}= a_{\mathrm{i}}/a_{\mathrm{eq}} > 1$ and $\alpha= a/a_{\mathrm{eq}} \gg 1$ which corresponds to the physically interesting situation where the initial conditions of the solution (\ref{SOLM3})  are set well after matter-radiation equality and anyway for redshifts smaller than $z_{\mathrm{drag}}$. In the latter limit  the 
result can be expressed as
\begin{equation}
\delta_{\mathrm{m}}^{(<)}(\vec{x},\alpha) = \frac{2}{3}  \biggl(\frac{\omega_{\mathrm{b}0}}{\omega_{\mathrm{M}0}} \biggr) 
\frac{(z_{\mathrm{eq}}+1)^3}{ H_{0}^2 \Omega_{\mathrm{M}0}} \frac{{\mathcal S}_{\mathrm{B}}(\vec{x},\alpha)}{\alpha^3}
+ {\mathcal D}_{1}(\vec{x}) \biggl( 1 + \frac{3}{2} \alpha\biggr)+ {\mathcal O}\biggl(\frac{1}{\alpha}\biggr) + {\mathcal O}\biggl(\frac{1}{\alpha_{\mathrm{i}}}\biggr),
\label{SR8}
\end{equation}
where, we remind, $\alpha= a/a_{\mathrm{eq}}$.  While in the regime 
$ a_{*} \leq a \ll a_{\mathrm{eq}}$ the total matter density contrast is well estimated by 
Eq. (\ref{SR8}),  for  $a > a_{*}$ the solution is instead given by Eq. (\ref{SOLM3}).
The initial conditions for the calculation of the growth rate can therefore be 
set by requiring that 
\begin{equation}
\delta^{(<)}_{\mathrm{m}}(\vec{x}, a_{*}/a_{\mathrm{eq}}) = \delta_{\mathrm{m}}(\vec{x},a_{*}/a_{\mathrm{de}}),\qquad 
\frac{\partial}{\partial a}  \delta^{(<)}_{\mathrm{m}}\biggl|_{a = a_{*}} =  \frac{\partial}{\partial a}  \delta_{\mathrm{m}}\biggl|_{a = a_{*}}.
\label{C1}
\end{equation}
The two conditions reported in Eq. (\ref{C1}) fix the constants appearing in Eq. (\ref{SOLM3}).
The growth rate \cite{gamma,growth} (see also \cite{bb0}) can be computed from Eq. (\ref{SOLM3}) and it is given by
\begin{equation}
f(\vec{x}, \alpha) = \frac{\partial \ln{\delta_{\mathrm{m}}}}{\partial\ln{\alpha}},
\label{GR1}
\end{equation}
where, according to Eq. (\ref{SOLM3}), 
$\delta_{\mathrm{m}}(\vec{x},\alpha)$ can be written as
\begin{equation}
\delta_{\mathrm{m}}(\vec{x},\alpha) = 
\delta(\vec{x}) \biggl[ {\mathcal C}_{1}(w_{\mathrm{de}}, z_{\mathrm{de}}, z_{*}) 
{\mathcal F}_{4}(\alpha,w_{\mathrm{de}}) + {\mathcal C}_{2}(w_{\mathrm{de}}, z_{\mathrm{de}}, z_{*}) 
{\mathcal F}_{5}(\alpha,w_{\mathrm{de}})\biggr] + \Sigma_{\mathrm{B}}(\vec{x}, \alpha).
\label{GR2}
\end{equation}
The two constants 
${\mathcal C}_{1}(w_{\mathrm{de}}, z_{\mathrm{de}}, z_{*})$ and ${\mathcal C}_{2}(w_{\mathrm{de}}, z_{\mathrm{de}}, z_{*})$ are a lengthy combination of hypergeometric 
functions depending on $a_{*} = 1/(z_{*} +1)$ and 
$a_{\mathrm{de}} = 1/(z_{\mathrm{de}} +1)$. The quantity $\delta(\vec{x})$ denotes the spatial 
profile of the fluctuation of the total density contrast which can be estimated in terms 
of the matter power spectrum while $\Sigma_{\mathrm{B}}(\vec{x},\alpha)$ denotes 
the contribution of the magnetic sources. The power spectrum of $\delta(\vec{x})$ can be written as 
\begin{eqnarray}
\langle \delta(\vec{x}) \, \delta(\vec{x} + \vec{r}) \rangle &=& 
\int d \ln{k} \, P_{\delta}(k) \, \frac{\sin{k\,r}}{k\, r}, 
\nonumber\\
{\mathcal P}_{\delta}(k,y_{\mathrm{eq}}) &=& \frac{4}{25}{\mathcal A}_{{\mathcal R}} \biggl(\frac{k}{k_{\mathrm{p}}}\biggr)^{n_{\mathrm{s}} -1} \ln^2{(k/k_{\mathrm{eq}})}, 
\label{GR3}
\end{eqnarray}
holding for wavenumbers larger than $k_{\mathrm{eq}} = 0.00974_{-0.00040}^{0.00041}\, \mathrm{Mpc}^{-1}$. In Eq. (\ref{GR3}) $k_{\mathrm{p}}=0.002\, \mathrm{Mpc}^{-1}$ and 
 ${\mathcal A}_{{\mathcal R}} = (2.43\pm 0.11)\times 10^{-9}$ in the case of  the WMAP 7yr data alone \cite{wmap7a,wmap7b,wmap7c} and for the case of the $\Lambda$CDM scenario.

Typical wavenumbers  $k \gg k_{\mathrm{eq}}$ crossed inside the Hubble volume before 
matter-radiation equality. Conversely the very large length-scales (relevant for the region of the Sachs-Wolfe 
plateau and for the integrated Sachs-Wolfe effect) fall into the complementary regime $k \ll k_{\mathrm{eq}}$.
The wavenumbers  touched by the present considerations range from $k_{\mathrm{min}} = 0.01 \, h_{0}\, \mathrm{Mpc}^{-1}$ to, approximately, 
$k_{\mathrm{max}} = 0.3 \,h_{0}\, \mathrm{Mpc}^{-1}$. The range $k_{\mathrm{min}} \leq k \leq k_{\mathrm{max}}$ 
includes also the scale at which the spectrum becomes nonlinear, i.e. $k_{\mathrm{nl}} \geq 0.09 \, h_{0} \, 
\mathrm{Mpc}^{-1}$.  The illustration of the analytical results 
will be given in terms of the following fiducial set of parameters determined on the basis 
of the WMAP 7yr data alone \cite{wmap7a,wmap7b,wmap7c}:
\begin{equation}
( \Omega_{\mathrm{b}0}, \, \Omega_{\mathrm{c}0}, \Omega_{\mathrm{de}0},\, h_{0},\,n_{\mathrm{s}},\, \epsilon_{\mathrm{re}}) \equiv (0.0449,\, 0.222,\, 0.734,\,0.710,\, 0.963,\,0.088),
\label{FL10}
\end{equation}
where $\epsilon_{\mathrm{re}}$ denotes the optical depth at reionization and $n_{\mathrm{s}}$ is the spectral index 
of curvature perturbations. The power spectra contributing to $\Sigma_{\mathrm{B}}(\vec{x},\alpha)$ can be solely expressed in  terms 
of $\Omega_{\mathrm{B}}(\vec{x},\alpha)$ and $\sigma_{\mathrm{B}}(\vec{x},\alpha)$ by recalling that 
\begin{equation}
\frac{\vec{\nabla} \cdot (\vec{J} \times \vec{B})}{4 a^4 \rho_{\gamma}} =4 \nabla^2 \sigma_{\mathrm{B}} - \nabla^2 \Omega_{\mathrm{B}},\qquad \Omega_{\mathrm{B}}(\vec{x}, \alpha) =\frac{B^2(\vec{x})}{8\, \pi \, a^4 \rho_{\gamma}}, \qquad \partial_{i} \partial_{j} \Pi^{ij}_{\mathrm{B}} = \frac{4}{3} \rho_{\gamma} \nabla^2 \sigma_{\mathrm{B}},
\label{GR4}
\end{equation}
where $\Pi^{ij}_{\mathrm{B}}$ denotes  the anisotropic stress of the magnetic fields. 
\begin{figure}[!ht]
\centering
\includegraphics[height=5cm]{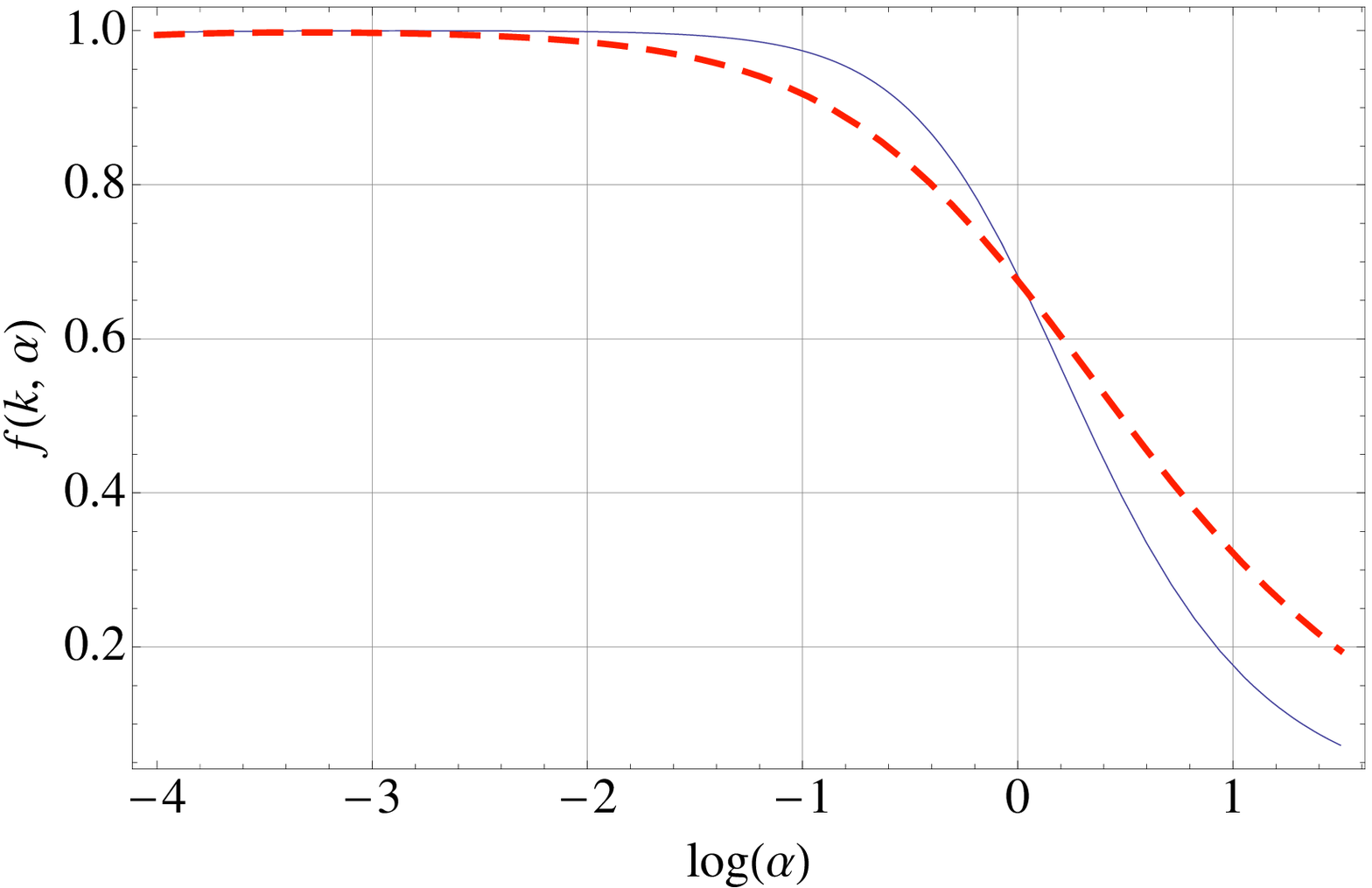}
\includegraphics[height=5cm]{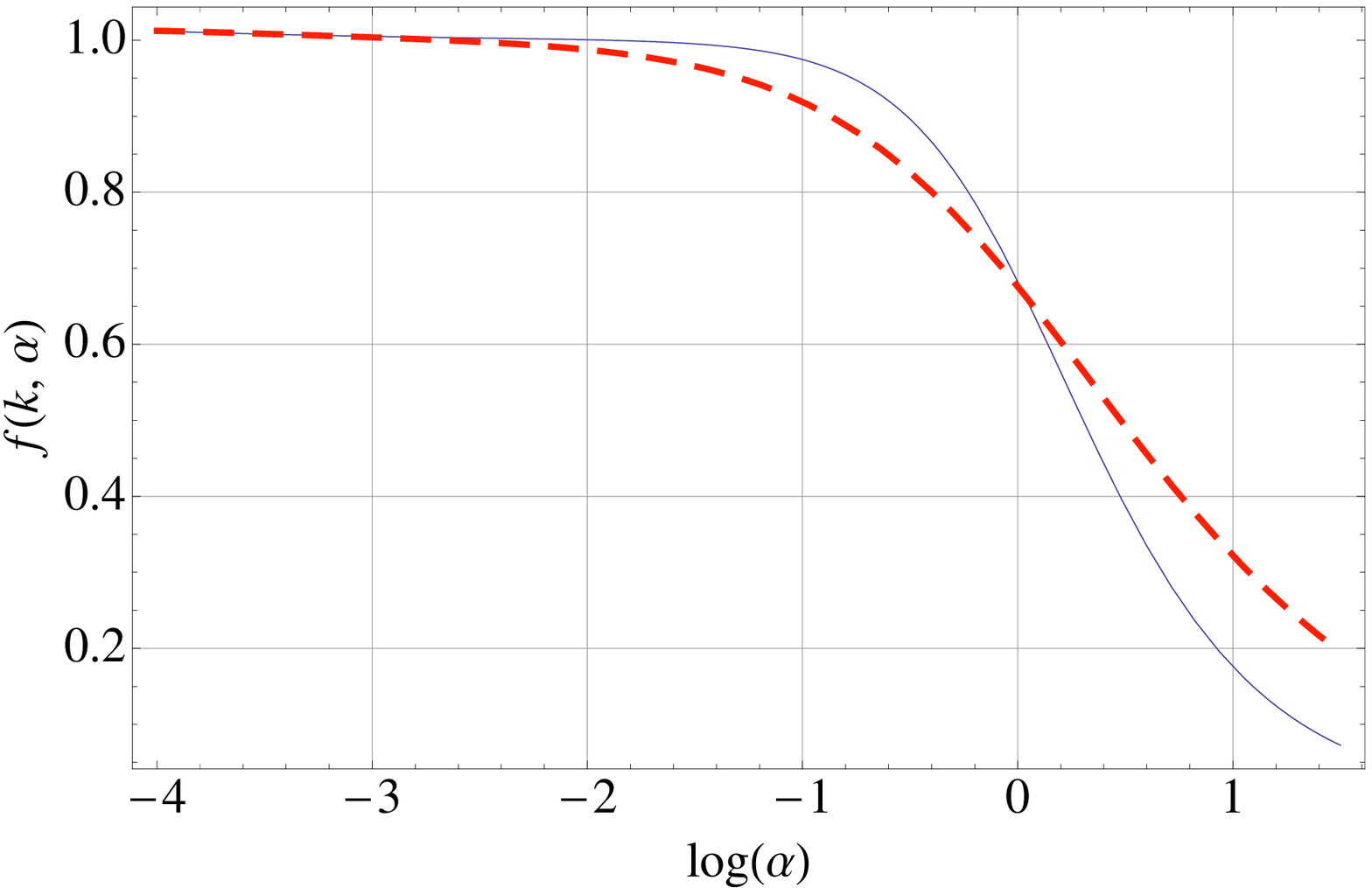}
\caption[a]{The growth rate is illustrated for two different values of the barotropic index, i.e. $w_{\mathrm{de}} = -1$ (full line) and $w_{\mathrm{de}} =-0.6$ (dashed line). In the plot at the 
left $B_{\mathrm{L}} =1 \, \mathrm{nG}$ and $k= 0.03\, \mathrm{Mpc}^{-1}$. In the plot at the right $B_{\mathrm{L}} =10 \, \mathrm{nG}$ and $k= 0.01\, \mathrm{Mpc}^{-1}$. In both plots the magnetic spectral index has been chosen as $n_{\mathrm{B}} = 1.5$.}
\label{F1}      
\end{figure}
In the range $k_{\mathrm{min}} \leq k \leq k_{\mathrm{max}}$, the terms containing the Ohmic current in the evolution equations (e.g. the term containing 
${\mathcal S}_{\mathrm{B}}(\vec{x},\alpha)$ in ${\mathcal T}^{(1)}_{\mathrm{B}}(\vec{x},\alpha)$) dominates against the one containing just $\Omega_{\mathrm{B}}(\vec{x},\tau)$ because of the presence of two supplementary spatial gradients. 
The same comment can be made for $\Sigma(\vec{x},\alpha)$.
In Fig. \ref{F1} the growth rate is reported for two illustrative values of the barotropic index 
$w_{\mathrm{de}} =-1$ (full line) and $w_{\mathrm{de}} = -0.6$ (dashed line); in both plots of 
Fig. \ref{F1}, $k = 0.03\, \mathrm{Mpc}^{-1}$. In Fourier space the power spectra of $\Omega_{\mathrm{B}}$ and $\sigma_{\mathrm{B}}$ 
will be denoted by ${\mathcal P}_{\Omega}(k)$ and ${\mathcal P}_{\sigma}(k)$  and their explicit expression 
can be written as
\begin{equation}
{\mathcal P}_{\Omega}(k) = {\mathcal E}_{\mathrm{B}} \biggl(\frac{k}{k_{\mathrm{L}}}\biggr)^{2 (n_{\mathrm{B}} -1) +\alpha_{\Omega}},\qquad {\mathcal P}_{\sigma}(k) 
= r_{\mathrm{B}} \,{\mathcal E}_{\mathrm{B}} \biggl(\frac{k}{k_{\mathrm{L}}}\biggr)^{2 (n_{\mathrm{B}} -1) +\alpha_{\sigma}},
\label{PS1}
\end{equation}
$\alpha_{\Omega}$ and $\alpha_{\sigma}$ are the corresponding running of the 
spectral indices which are set to zero in the minimal magnetized $\Lambda$CDM scenario.
The relative amplitude of the two power 
spectra at the magnetic pivot scale $k_{\mathrm{L}}$ (conventionally chosen to be $1\,\, \mathrm{Mpc}^{-1}$) is controlled by $r_{\mathrm{B}}$. Since the magnetic fields are, themselves, stochastically distributed,  the ensemble average of their Fourier modes obeys
\begin{equation}
\langle B_{i}(\vec{k}) \, B_{j}(\vec{p}) \rangle = \frac{2\pi^2 }{k^3} P_{ij}(k) {\mathcal P}_{\mathrm{B}}(k) \delta^{(3)}(\vec{k} + \vec{p}), \qquad {\mathcal P}_{{\mathrm{B}}}(k) = {\mathcal A}_{\mathrm{B}} \biggl(\frac{k}{k_{\mathrm{L}}}\biggr)^{n_{\mathrm{B}}-1}, 
 \label{PS2}
 \end{equation}
where $P_{ij}(k) = (k^2 \delta_{ij} - k_{i} k_{j})/k^2$ and  ${\mathcal A}_{\mathrm{B}}$ the spectral amplitude of the magnetic field. 
The spectral amplitude of the magnetic 
energy density ${\mathcal E}_{\mathrm{B}}$ depends upon the 
ultraviolet cut-off associated with diffusion damping (i.e. $k_{\mathrm{D}}$), upon 
the infrared cut-off associate with the (comoving) angular diameter distance 
at last scattering (i.e. $D_{\mathrm{A}}(z_{1})$) and also on
 the magnetic spectral index $n_{\mathrm{B}}$, i.e. \cite{mg1,mg2}
\begin{eqnarray}
&& r_{\mathrm{B}} = \frac{n_{\mathrm{B}} + 29}{20 (7 - n_{\mathrm{B}})}\biggl[ 1 + 
\frac{(5 - 2 n_{\mathrm{B}})(83 n_{\mathrm{B}}- 473)}{2 ( 7 - n_{\mathrm{B}}) 
(n_{\mathrm{B}} + 29)} \biggl(\frac{k}{k_{\mathrm{A}}}\biggr)^{1 - n_{\mathrm{B}}}\biggr],
\qquad n_{\mathrm{B}} < 1,
\label{PS3}\\
&& r_{\mathrm{B}}  = \frac{n_{\mathrm{B}} + 29}{20 (7 - n_{\mathrm{B}})}\biggl[1 + \frac{(n_{\mathrm{B}} -1)(87 n_{\mathrm{B}} -501)}{(n_{\mathrm{B}} + 29) ( 7 - n_{\mathrm{B}})}\biggl(\frac{k}{k_{\mathrm{D}}}\biggr)^{5 - 2 n_{\mathrm{B}}}\biggr] ,\qquad
n_{\mathrm{B}} >1,
\label{PS4}
\end{eqnarray}
where, recalling that $D_{A}(z_{1}) = 2 d_{\mathrm{A}}(z_{1})/(H_{0} \sqrt{\Omega_{\mathrm{M}0}})$, 
\begin{equation}
k_{\mathrm{A}}(z_{1}) = 1/D_{\mathrm{A}}(z_{1}),\qquad \frac{k_{\mathrm{D}}(z_{1})}{k_{\mathrm{A}}(z_{1})}= \frac{2240 \, d_{\mathrm{A}}(z_{1})}{\sqrt{\sqrt{r_{\mathrm{R}1} +1} - \sqrt{r_{\mathrm{R}1}}}} 
\biggl(\frac{z_{1}}{10^{3}} \biggr)^{5/4} \, \omega_{\mathrm{b}}^{0.24} \omega_{\mathrm{M}}^{-0.11}.
\label{PS4a}
\end{equation}
 In Eq. (\ref{PS4a}) $r_{\mathrm{R}1}$ is
the ratio of radiation to matter at $z_{1}$ which is the redshift 
of last-scattering and which can be determined analytically for typical values 
of the parameters close to the best-fit determined on the basis of the WMAP 7yr data 
and in the light of the $\Lambda$CDM paradigm:
\begin{eqnarray}
 && r_{\mathrm{R}1} = \frac{\rho_{\mathrm{R}}(z_{1})}{\rho_{\mathrm{M}}(z_{1})} =
\frac{ 4.15\times 10^{-2}}{\omega_{\mathrm{M}}}\, \biggl(\frac{z_{1}}{10^{3}}\biggr),
 \qquad 
z_{1} = 1048[ 1 + f_{1} \omega_{\mathrm{b}}^{- f_{2}}] [ 1 + g_{1} \omega_{\mathrm{M}}^{g_2}],
\nonumber\\
&& g_{1} = \frac{0.0783 (\omega_{\mathrm{b}})^{-0.238}}{[1 + 39.5 (\omega_{\mathrm{b}})^{0.763}]},\qquad g_{2} = \frac{0.560}{1 + 21.1 (\omega_{\mathrm{b}})^{1.81}},
\label{PS4b}
\end{eqnarray}
where $f_{1}$ and $f_{2}$ are given, respectively, by $f_{1} = 1.24\times 10^{-3}$,
$f_{2} = 0.738$.  When $n_{\mathrm{B}}=1$ the spectra for the energy density and for the Lorentz force are scale-invariant once the logarithmic divergence of the two-point function 
is appropriately subtracted.  The amplitude ${\mathcal A}_{\mathrm{B}}$ can be traded 
for the magnetic field regularized over a spatial domain $k_{\mathrm{L}}^{-1}$, i.e. 
in the case $n_{\mathrm{B}} > 1$ 
\begin{equation}
{\mathcal E}_{\mathrm{B}} = \frac{4 ( 7 - n_{\mathrm{B}})}{3 (n_{\mathrm{B}} -1) (5 - 2 n_{\mathrm{B}})} \frac{{\mathcal A}_{\mathrm{B}}^2}{(8\pi \overline{\rho}_{\gamma})^2},
\qquad {\mathcal A}_{\mathrm{B}} = \frac{(2\pi)^{n_{\mathrm{B}} -1}}{\Gamma[(n _{\mathrm{B}}-1)/2]} B_{\mathrm{L}}^2,
\label{PS5}
\end{equation}
and analogously in the $n_{\mathrm{B}} <1$ case but with ${\mathcal A}_{\mathrm{B}} =[ (1 -n_{\mathrm{B}})/2] (k_{\mathrm{A}}/k_{\mathrm{L}})^{(1 - n_{\mathrm{B}})}B_{\mathrm{L}}^2$. 

In the case of the minimal m$\Lambda$CDM the relation 
of $(n_{\mathrm{B}}, {\mathcal E}_{\mathrm{B}})$ to  
$(n_{\mathrm{B}}, B_{\mathrm{L}})$ follows from Eqs. (\ref{PS2})--(\ref{PS5}). 
The magnetic energy density can be naturally referred to the energy density 
of the photon background so that $\Omega_{\mathrm{B\,L}} = B_{\mathrm{L}}^2/(8\pi \overline{\rho}_{\gamma})$ can be measured in units of the amplitude 
of curvature perturbations:
\begin{equation} 
\frac{\overline{\Omega}_{\mathrm{B\,L}}}{{\mathcal A}_{\mathcal R}}=  39.568 \biggl(\frac{B_{\mathrm{L}}}{\mathrm{nG}}\biggr)^{2} \, \biggl(\frac{T_{\gamma0}}{2.725\,\mathrm{K}}\biggr)^{-4} \, \biggl(\frac{{\mathcal A}_{{\mathcal R}}}{2.41\times 10^{-9}}\biggr)^{-1}.
\label{PS6}
\end{equation}
The parameter space  of the magnetized $w$CDM models and of the magnetized  $\Lambda$CDM scenario 
have been investigated, respectively, in \cite{mg1} and \cite{mg2}. In a frequentist approach, the boundaries of the confidence regions obtained in \cite{mg1,mg2} represent exclusion plots at  $68.3\,$\% and $95.4\,$\% confidence level.  
When moving from the magnetized $\Lambda$CDM 
scenario to the magnetized $w$CDM model we have that the 
 the parameters maximizing the likelihood get shifted to slightly larger values\footnote{The difference between Eqs. (\ref{NTT}) and (\ref{NTE}) is that the parameters 
of Eq. (\ref{NTT}) are obtained from the analysis of the temperature autocorrelations while the parameters of Eq. (\ref{NTE})  are obtained 
by adding the data points of the cross-correlations between temperature and E-mode polarization \cite{mg1}.}
\begin{eqnarray}
&&(n_{\mathrm{B}},B_{\mathrm{L}})_{\Lambda\mathrm{CDM}} = ( 1.598,\,3.156 \mathrm{nG})\to (n_{\mathrm{B}} , B_{\mathrm{L}})_{\mathrm{{\it w}CDM}} =(1.883, \,4.982\, \mathrm{nG}), 
\label{NTT}\\
&&(n_{\mathrm{B}}, B_{\mathrm{L}})_{\Lambda\mathrm{CDM}} = ( 1.616,\,3.218 \mathrm{nG})\to (n_{\mathrm{B}},B_{\mathrm{L}})_{\mathrm{{\it w}CDM}} = (1.913, \,5.163\, \mathrm{nG}).
\label{NTE}
\end{eqnarray}
Even if the addition of a fluctuating dark energy background pins down systematically larger values of the magnetic field  parameters, the results of \cite{mg1,mg2} will be used here just for a consistent illustration
of the results. 
\begin{figure}[!ht]
\centering
\includegraphics[height=5cm]{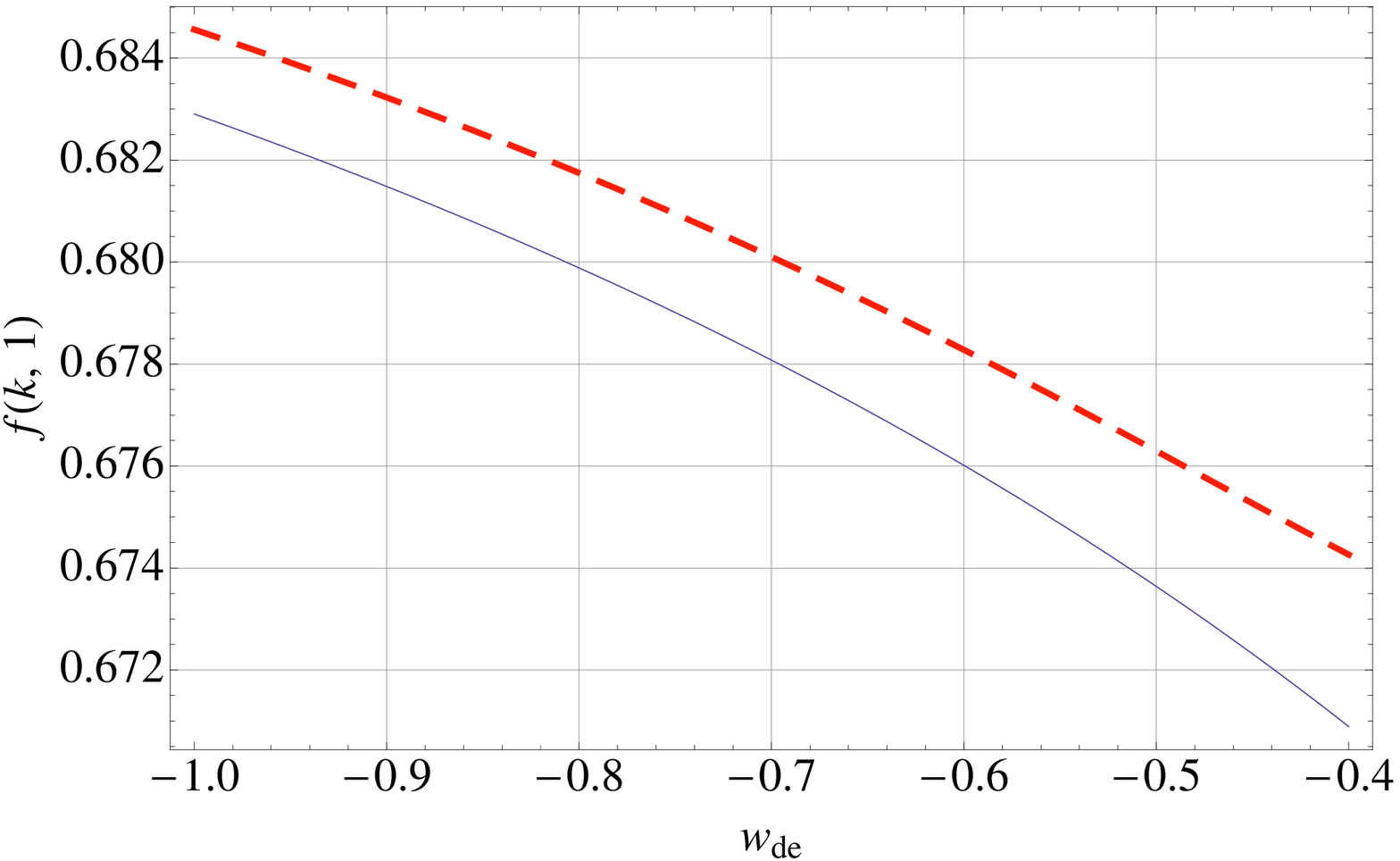}
\includegraphics[height=5cm]{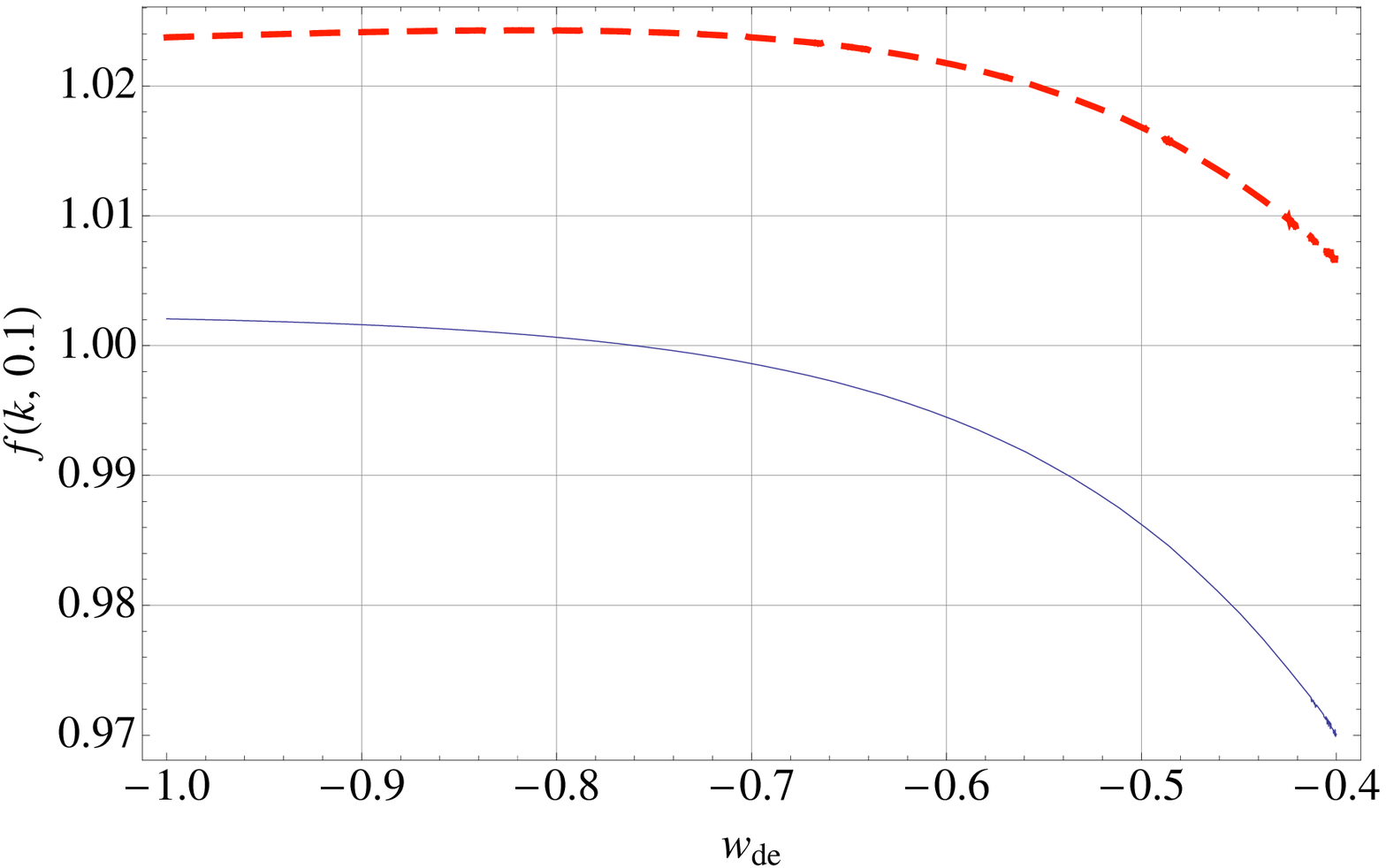}
\caption[a]{In the plot at the left the growth rate is illustrated for $\alpha =1$ (i.e. $a= a_{\mathrm{de}}$)
for $B_{\mathrm{L}} = 1\, \mathrm{nG}$ (full line) and $B_{\mathrm{L}} = 10 \, \mathrm{nG}$ (dashed line).
In the right plot $\alpha =0.1$ (i.e. $a = 0.1 \, a_{\mathrm{de}}$) and the values of $B_{\mathrm{L}}$ are, respectively, 
   $B_{\mathrm{L}} = 10\, \mathrm{nG}$ (full line) and $B_{\mathrm{L}} = 30\, \mathrm{nG}$ (dashed line). In both plots $n_{\mathrm{B}} =1.5.$ and $k =0.03\, \mathrm{Mpc}^{-1}$.}
\label{F2}      
\end{figure}
In Fig. \ref{F2} the growth rate is illustrated as a function of $w_{\mathrm{de}}$ 
and for two different choices of $B_{\mathrm{L}}$, i.e. $1\, \mathrm{nG}$ (full line)
and $10\, \mathrm{nG}$ (dashed line).  The difference between the plot at the left and at the right is the value of the redshift: while in the plot at the left $a= a_{\mathrm{de}}$ (i.e. $z = z_{\mathrm{de}}$)  in the plot at the right $a= 0.1\,a_{\mathrm{de}}$ and $z > z_{\mathrm{de}}$.
\begin{figure}[!ht]
\centering
\includegraphics[height=5cm]{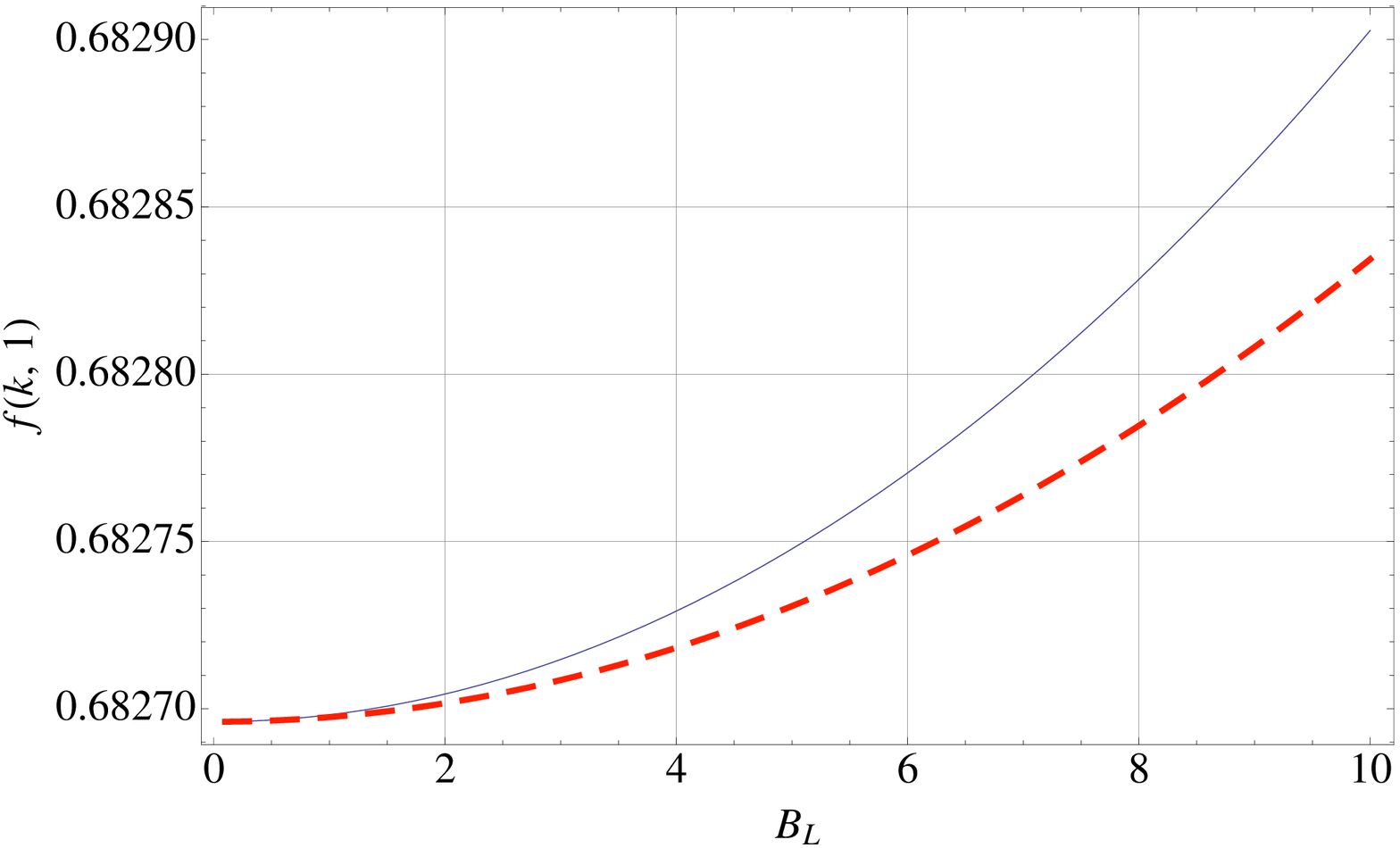}
\includegraphics[height=5cm]{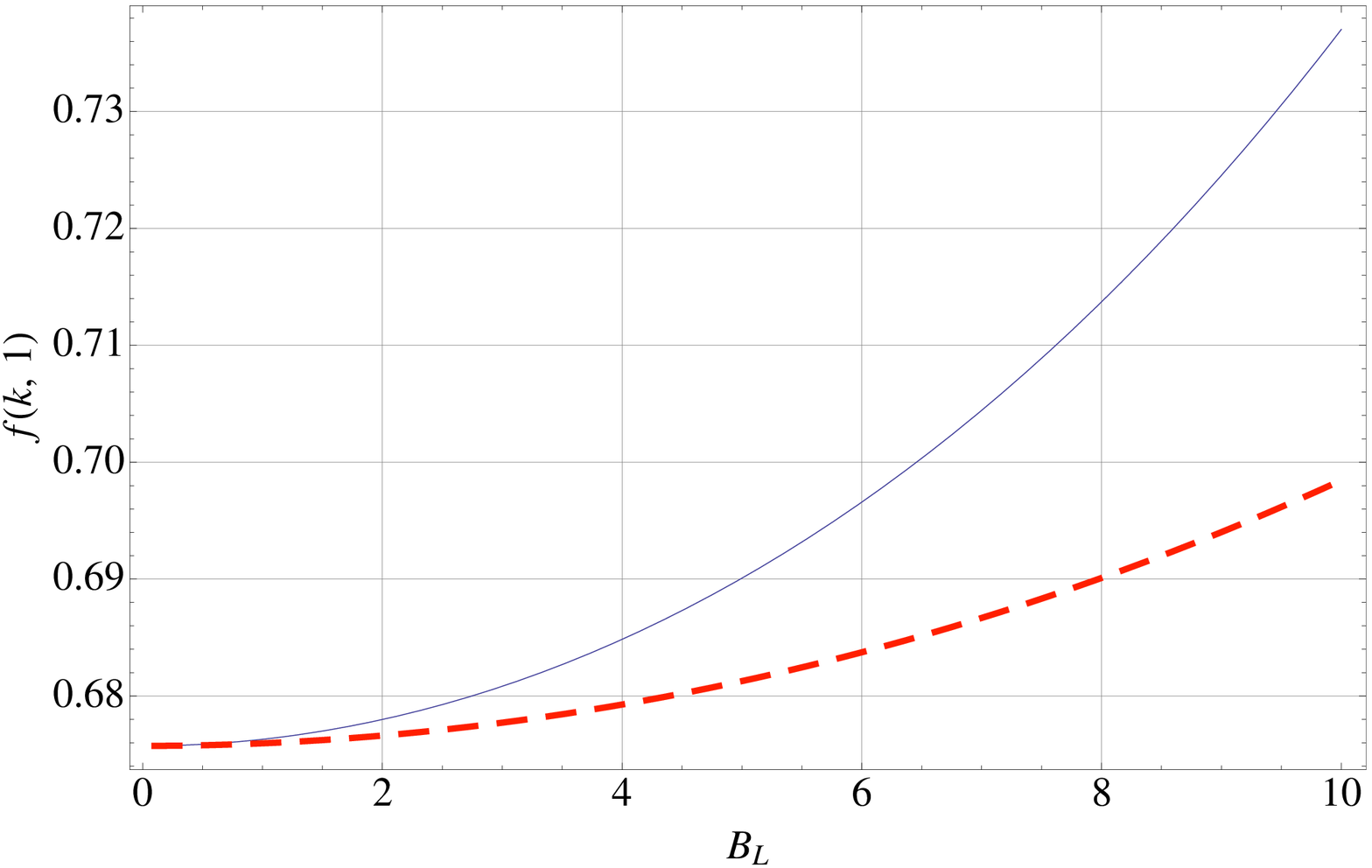}
\caption[a]{In both plots the growth rate is illustrated for $a= a_{\mathrm{eq}}$ as a function of the 
regularized magnetic field intensity $B_{\mathrm{L}}$. In the plot at the left $w_{\mathrm{de}} = -1$ (and $k =0.03\, \mathrm{Mpc}^{-1}$) while in the plot at the right $w_{\mathrm{de}} = -0.6$ (and $k=0.3\,\mathrm{Mpc}^{-1}$).}
\label{F3}      
\end{figure}
In Fig. \ref{F3} the growth rate is illustrated as a function of $B_{\mathrm{L}}$ 
for different choices of the wavenumber and of the barotropic index. The 
full and dashed lines correspond, in each plot of Fig. \ref{F3}, to 
two different values of the magnetic spectral index, i.e. $n_{\mathrm{B}} = 1.5$ 
(full line) and $n_{\mathrm{B}} =1.2$ dashed line.

\begin{figure}[!ht]
\centering
\includegraphics[height=5cm]{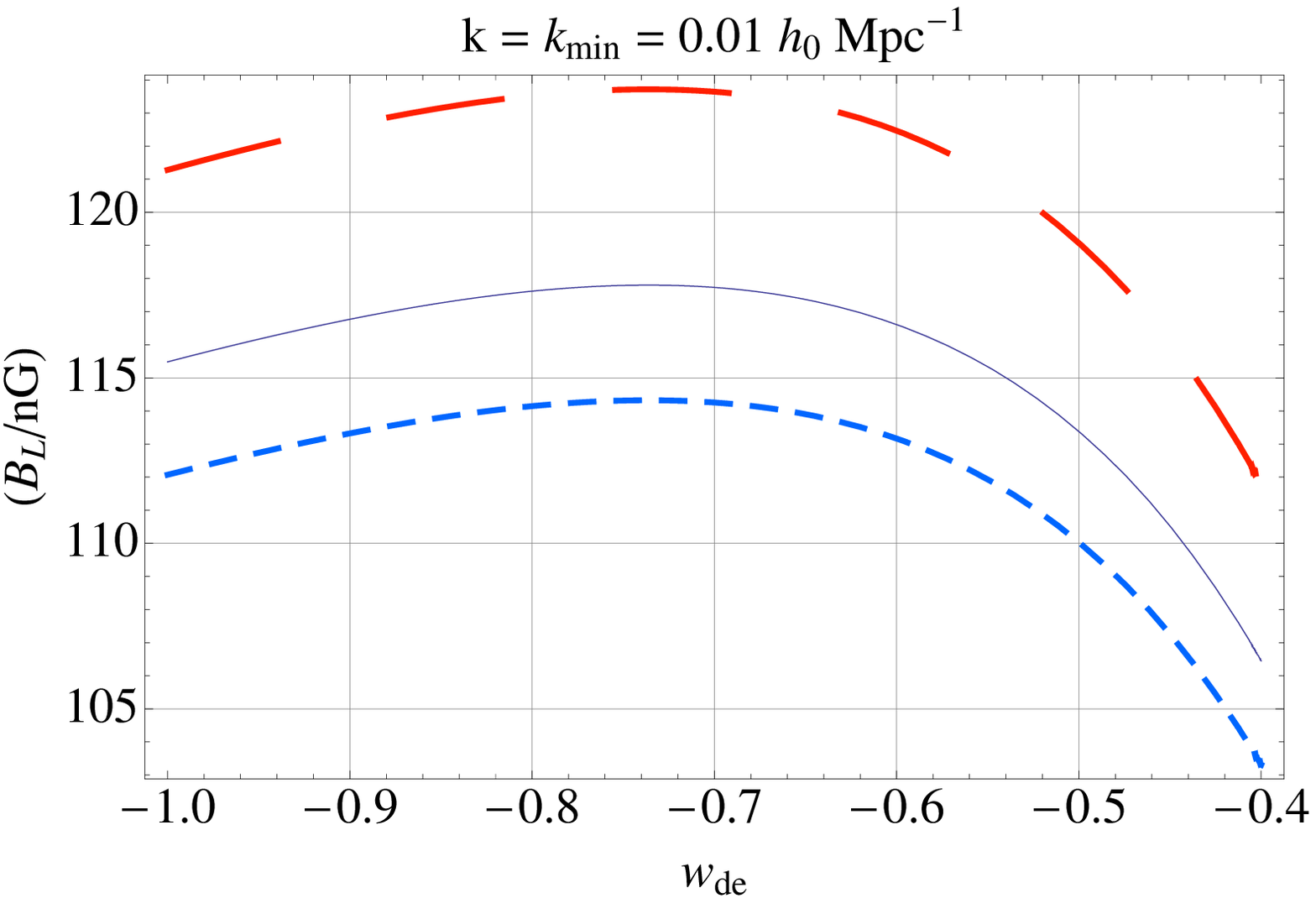}
\includegraphics[height=5cm]{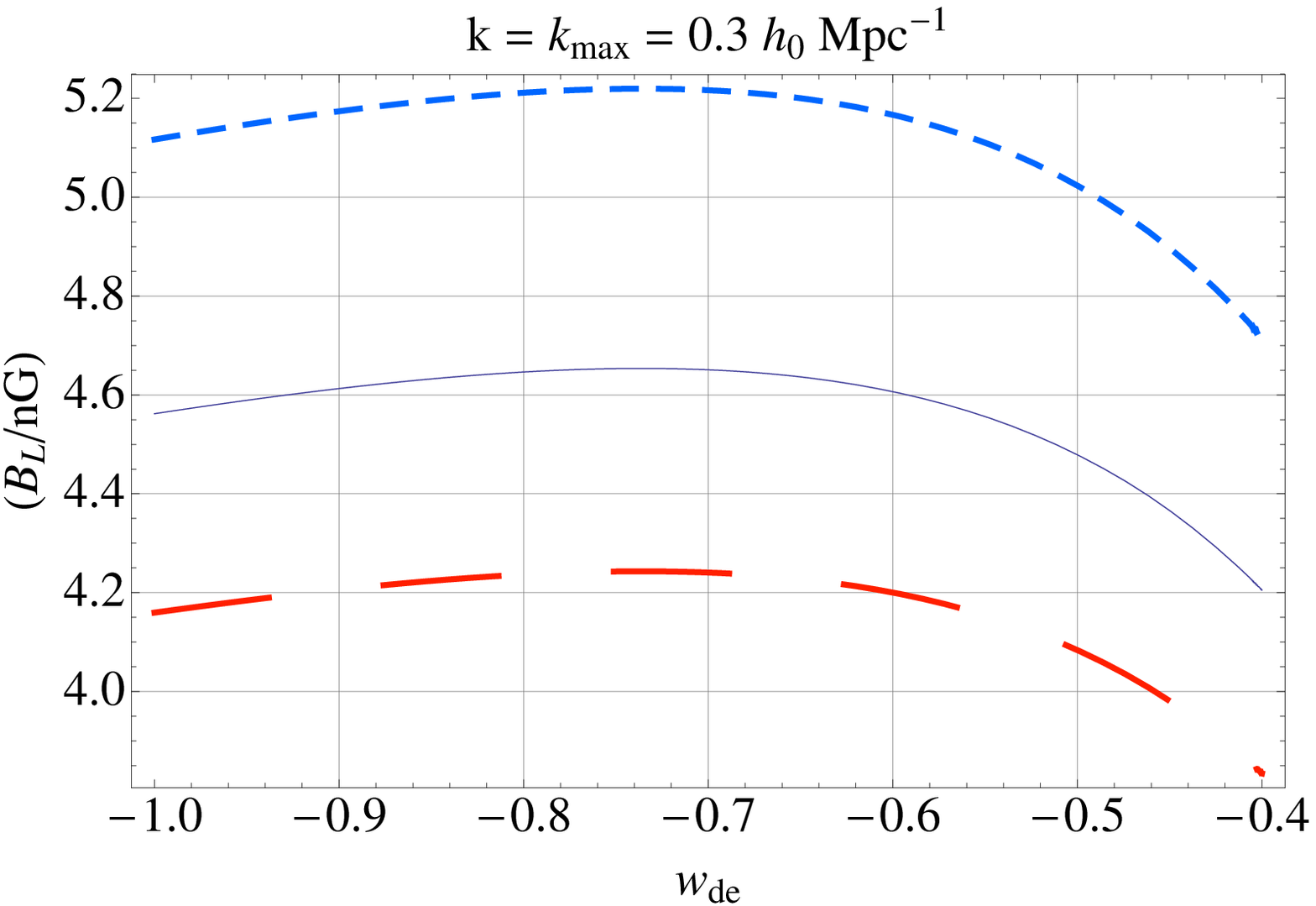}
\caption[a]{The bound on $B_{\mathrm{L}}$ illustrated 
for $n_{\mathrm{B}} = 1.6$ (long dashed line), $n_{\mathrm{B}} = 1.5$ (full line) 
and $n_{\mathrm{B}} = 1.4$ (short dashed line). In the left plot 
$k=0.03\, \mathrm{Mpc}^{-1}$. In the right plot $k=0.3\, \mathrm{Mpc}^{-1}$. In both plots $a = a_{\mathrm{eq}}$}
\label{F4}      
\end{figure}

In the 
absence of large-scale magnetic fields, the growth rate can be parametrized as $\Omega_{\mathrm{M}}^{\gamma}(\alpha)$ where $\gamma$ is the growth index which can be 
explicitly estimated as \cite{gong} (see also \cite{gamma})
\begin{equation}
\gamma(w_{\mathrm{de}}) =\frac{ 3 w_{\mathrm{de}} - 3}{6 w_{\mathrm{de}} - 5} 
+ \frac{3}{125} \frac{(1- w_{\mathrm{de}})(1 - 3 w_{\mathrm{de}}/2)}{(1- 6 w_{\mathrm{de}}/5)^2}\epsilon +  {\mathcal O}(\epsilon^2),
\label{BB1}
\end{equation}
where $\epsilon = 1 - \Omega_{\mathrm{M}}$. The requirement  that 
the magnetized growth rate of Eq. (\ref{GR1}) does not exceed 
the standard fit for the growth rate implies an upper bound 
on $B_{\mathrm{L}}$ which is illustrated in Fig. \ref{F4}.
The bound on $B_{\mathrm{L}}$ gets more stringent as $k$ increases and for 
large redshift. In Fig. \ref{F4} $a= a_{\mathrm{de}}$. The allowed values of $B_{\mathrm{L}}$ for 
different values of $w_{\mathrm{de}}$ stay below each of the curves reported in the plots of Fig. \ref{F4}.
Recalling the definition of the comoving magnetic Jeans length \cite{bb0}
\begin{equation}
\lambda_{\mathrm{B\,J}} = c_{\mathrm{a}} \sqrt{\frac{\pi}{G \, a^3 \rho_{\mathrm{b}}}}=1.90\times 10^{-2} \, \biggl(\frac{\omega_{\mathrm{b}0}}{0.02258}\biggr)^{-1} \, 
\biggl(\frac{B_{\mathrm{L}}}{\mathrm{nG}}\biggr)\,\,\mathrm{Mpc}, \qquad 
c_{\mathrm{a}}^2 = \frac{B_{\mathrm{L}}^2}{8 \pi a^3 \,\rho_{\mathrm{b}}},
\label{MJ1}
\end{equation}
the results of Fig. \ref{F4} imply $\lambda_{\mathrm{B\, J}}  \leq {\mathcal O}(\mathrm{Mpc})$.
The considerations reported in the present analysis suggest that the study of large-scale magnetism 
during the dark ages provide further constraints on the evolution of large-scale magnetic fields. The 
quantitative features of the obtained constraints have been shown to be both complementary and competitive 
 with the bounds stemming from a direct analysis of potential distortions induced on the temperature and polarization anisotropies for a range of length scales shorter than the one relevant for the physics of the CMB \cite{rev}. 
 
If the origin of the large-scale magnetic fields is primordial (as opposed to astrophysical) it is plausible to expect the presence of magnetic fields in the primeval plasma also  before the decoupling of radiation from matter.  If the evolution of the large-scale magnetic fields follows the tenets of magnetohydrodynamics then the magnetic flux will be conserved across the last scattering and potential effects of the magnetic fields during the dark ages could help to decide on their origin and implications.   In this perspective the SKA 
\cite{SKA} program could  provide decisive informations both from 
the foreseen full sky surveys of Faraday rotation and  also from more detailed pictures of structure formation and reionization over a wide range of redshifts 
during the dark age.


\newpage 

\begin{thebibliography}{99}

\bibitem{wmap7a} C.~L.~Bennett {\it et al.},  Astrophys.\ J.\ Suppl.\  {\bf 192}, 17 (2011);   
N.~Jarosik {\it et al.},  Astrophys.\ J.\ Suppl.\  {\bf 192}, 14 (2011).

\bibitem{wmap7b} J.~L.~Weiland {\it et al.},  Astrophys.\ J.\ Suppl.\  {\bf 192}, 19 (2011); 
D.~Larson {\it et al.}, Astrophys.\ J.\ Suppl.\  {\bf 192}, 16 (2011).  

\bibitem{wmap7c} B.~Gold {\it et al.},  Astrophys.\ J.\ Suppl.\  {\bf 192}, 15 (2011);  
E.~Komatsu {\it et al.},   Astrophys.\ J.\ Suppl.\  {\bf 192}, 18 (2011).

\bibitem{sdss1}   L.~-H.~Jiang,  {\it et al.},
  Astron.\ J.\  {\bf 135}, 1057-1066 (2008);  X.~-H.~Fan, 
  Astron.\ J.\  {\bf 132}, 117 (2006).

\bibitem{sdss2} J.~K.~Adelman-McCarthy {\it et al.} [ SDSS Collaboration ],  Astrophys.\ J.\ Suppl.\  {\bf 175}, 297 (2008);
  K.~N.~Abazajian {\it et al.} [ SDSS Collaboration ],   Astrophys.\ J.\ Suppl.\  {\bf 182}, 543 (2009).
  
\bibitem{mg1} M.~Giovannini, Phys.\ Rev.\  {\bf D79}, 121302 (2009);  Phys.\ Rev.\  {\bf D79}, 103007 (2009).

\bibitem{mg2} M. Giovannini,  Class.\ Quant.\ Grav.\  {\bf 27}, 105011 (2010).

\bibitem{lofar} See http://www.lofar.org/  for more informations.

\bibitem{mwa} See http://www.haystack.mit.edu/ast/arrays/mwa/ for more informations.

\bibitem{VLA} See http://www.vla.nrao.edu/ for more informations.

\bibitem{SKA} See http://www.skatelescope.org for more informations.

\bibitem{hwang1}  J. Hwang, Astrophys. J.  {\bf 375}, 443 (1990); J.~Hwang,  Astrophys.\ J.\  {\bf 427}, 533 (1994).

\bibitem{hwang2}  J.~-C.~Hwang, H.~Noh,  Phys.\ Rev.\  {\bf D54}, 1460 (1996); J.~Hwang and H.~Noh,   Class.\ Quant.\ Grav.\  {\bf 19}, 527 (2002).

\bibitem{et1} V.~Silveira, I.~Waga, Phys.\ Rev.\  {\bf D56}, 4625(1997).
  
\bibitem{et2} D. J. Heath, Mon. Not. R. Astr. Soc. {\bf 179}, 351 (1977).  

\bibitem{et3} P.~Meszaros,   Astron.\ Astrophys.\  {\bf 37 },  225 (1974).

\bibitem{drag1} D.~J.~Eisenstein and W. Hu, Astrophys.\ J.\  {\bf 496}, 605 (1998).

\bibitem{abr} M. Abramowitz and I. A. Stegun, {\it Handbook of Mathematical Functions} (Dover, New York, 1972).

\bibitem{tric}  A. Erdelyi, W. Magnus, F. Obehettinger, and F. Tricomi,  {\it Higher Trascendental Functions} (Mc Graw-Hill, New York, 1953).  

\bibitem{PV1} W. Press and E. Vishniac, Astrophys. J. {\bf 239}, 1 (1980);
 Astrophys. J. {\bf 236}, 323 (1980).
 
 \bibitem{cosmics2}   C.~P.~Ma and E.~Bertschinger,  Astrophys.\ J.\  {\bf 455}, 7 (1995)

\bibitem{bardeen}  J. M. Bardeen, Phys. Rev. D {\bf 22}, 1882 (1980).

\bibitem{weinberg}  S.~Weinberg,  Phys.\ Rev.\  D {\bf 67}, 123504 (2003).

\bibitem{cs1} R.~Bean and O.~Dore,  Phys.\ Rev.\  D {\bf 69}, 083503 (2004).

\bibitem{cs2} H.~Kodama and M.~Sasaki,  Prog.\ Theor.\ Phys.\ Suppl.\  {\bf 78}, 1 (1984).

\bibitem{gamma} P.J.E. Peebles,  Astrophys. J. {\bf 205}, 318 (1976);  L. Wang, P. J. Steinhardt,  Astrophys. J. {\bf 508}, 483 (1998). 

\bibitem{nist} F. W. J. Olver, D. W. Lozier, R.  F. Boisvert and C. W. Clark, {\it NIST handbook of mathematical functions}, (Cambridge University Press, Cambridge).

\bibitem{growth}  E.~Bertschinger,  Astrophys.\ J.\  {\bf 648}, 797 (2006).
 
\bibitem{bb0} P. J. E. Peebles, {\it The large-scale structure of the Universe}, (Princeton University Press, Princeton, NJ).

\bibitem{gong} Y.~Gong, M.~Ishak, A.~Wang,  Phys.\ Rev.\  {\bf D80}, 023002 (2009).

\bibitem{bb1} I. Wasserman,  Astrophys. J. {\bf 224}, 337 (1978); P. Coles, Comments Astrophys. {\bf 16}, 45 (1992).

\bibitem{rev} M.~Giovannini,  Class.\ Quant.\ Grav.\  {\bf 23}, R1 (2006).


\end{thebibliography}
\end{document}